\documentclass[11pt,a4paper]{article} 
\pdfoutput=1

\usepackage[no-natbib-sort]{my-jheppub}

\usepackage{amsmath, amssymb}
\usepackage{mathpazo}
\usepackage{environ}
\usepackage{mathrsfs}
\usepackage{array,arydshln}

\usepackage{graphicx,epsfig}
\usepackage{epic}
\usepackage{youngtab}
\usepackage{float}
\usepackage{color}
\definecolor{maroon}{rgb}{0.8,0.3,0.}

\usepackage{slashed}

\usepackage[nodayofweek]{datetime}

\usepackage{aurical}
\usepackage[T1]{fontenc}

\newcommand{\be}{\begin{equation}}
\newcommand{\ee}{\end{equation}}

\newcommand{\ads}{AdS$_5\times S^5$\ }

\newcommand{\mc}{\mathcal }
\newcommand{\Z}{\mathcal{Z}}

\def\XXint#1#2#3{{\setbox0=\hbox{$#1{#2#3}{\int}$}
     \vcenter{\hbox{$#2#3$}}\kern-.5\wd0}}

\def \del{ \partial}
\def \la {\label}
\newcommand{\rf}[1]{(\ref{#1})}
\def\ov{\over}
\def\no{\nonumber} \def \aa {{\rm a}}

\def \ci {\cite}
\def \p {\phi}
\def \m {\mu}\def \n {\nu} 
\def \ed {\end{document}}
\def \l {\lambda} \def \r {\rho} 

\def \foot {\footnote}
\def \b {\beta} 
\def \dd {{\rm d}} 
\def \om {\omega} 
\def \tr {{\rm tr}}
\def \D {\Delta} 
\def \vp {\varphi} 
 \def \ha {{{1 \ov 2}}}

\def \cc {{\rm c}}
\def \aa {{\rm a} }

\def \OO  {{\mc O}}   \def \tr  {{\rm tr }}   
\def \g {\gamma} 

\def \CHS {{\rm CHS}}
\def \MHS {{\rm MHS}}

\def \bDelta {{\bf \Delta}}
 \def \PP {{\rm P}}   \def \FF {{\rm F}} 
\def \hDelta {{\hat \Delta}}
\def \L  {\Lambda} 
 \def \GG {{\rm G}}

\def \LL {{\rm L}}

\def \iffa  {\iffalse}

\def \iffa  {\iffalse}

\title{On     higher spin  partition functions  }   

\author[a]{Matteo Beccaria} 
\author[b]{  and \   Arkady A. Tseytlin\footnote{Also at Lebedev Institute, Moscow}}

\affiliation[a]{Dipartimento di Matematica e Fisica Ennio De Giorgi,\\
Universit\`a del Salento \& INFN, Via Arnesano, 73100 Lecce, 
Italy} 

\affiliation[b]{Blackett Laboratory, Imperial College, London SW7 2AZ, U.K.}

\emailAdd{matteo.beccaria@le.infn.it}
\emailAdd{tseytlin@imperial.ac.uk}

\abstract{
We  observe  that the  partition function  of the set of  all  free  massless higher spins  $s=0,1,2,3,...   $
in flat space  is equal to one:   the ghost determinants   cancel against the "physical" ones  
or, equivalently, the (regularized) total  number of degrees of freedom vanishes. 
This reflects  large underlying gauge symmetry and   suggests  analogy with  supersymmetric or topological theory. 
The   $Z=1 $  property  extends  also to  the AdS   background, i.e. the  
1-loop   vacuum partition function  of  Vasiliev theory is  equal to 1   (assuming a particular regularization of the sum over spins);
this  was   noticed   earlier  as  a  consistency requirement for  the vectorial AdS/CFT duality.
We  find that $Z=1$  is true also  in the   conformal higher spin theory  (with higher-derivative  $\del^{2s} $  kinetic terms)  
expanded near  flat or  conformally flat  $S^4$  background. 
 We also  consider  the  partition function of   free   conformal   theory of symmetric traceless rank $s$  tensor field  which has   2-derivative  kinetic term
  but  only scalar gauge invariance in  flat  4d  space. This non-unitary theory    has Weyl-invariant    action in    curved   background   and 
   it  corresponds 
 to   "partially massless"  field in $AdS_5$. We discuss  in detail  the special case of  $s=2$  (or "conformal graviton"),  compute the corresponding 
  conformal anomaly coefficients   and compare them with   previously found  expressions  for   generic  representations  of  conformal group in 4 dimensions. 
  }

    \iffa 
Authors:
M. Beccaria,  A.A. Tseytlin
Report No.: Imperial-TP-AT-2015-01
Title:     
Comments:    34 pages. 
Abstract: 

We  observe  that the  partition function  of the set of  all  free  massless higher spins  s=0,1,2,3,...   
in flat space  is equal to one:   the ghost determinants   cancel against the "physical" ones  
or, equivalently, the (regularized) total  number of degrees of freedom vanishes. 
This reflects  large underlying gauge symmetry and   suggests an analogy with  supersymmetric or topological theory. 
The   Z=1   property  extends  also to  the AdS   background, i.e. the  
1-loop   vacuum partition function  of  Vasiliev theory is  equal to 1   (assuming a particular regularization of the sum over spins);
this  was   noticed   earlier  as  a  consistency requirement for  the vectorial AdS/CFT duality.
We  find that Z=1  is true also  in the   conformal higher spin theory  (with higher-derivative  d^{2s} kinetic terms)  
expanded near  flat or  conformally flat  S^4  background. 
We also  consider  the  partition function of   free   conformal   theory of symmetric traceless rank s  tensor field  which has   2-derivative  kinetic term
  but  only scalar gauge invariance in  flat  4d  space. This non-unitary theory    has Weyl-invariant    action in    curved   background   and 
   is  corresponds 
 to   "partially massless"  field in AdS_5. We discuss  in detail  the special case of  s=2  (or "conformal graviton"),  compute the corresponding 
  conformal anomaly coefficients   and compare them with   previously found  expressions  for   generic  representations  of  conformal group in 4 dimensions. 
 \fi

\allowdisplaybreaks

\begin{document}


\begin{flushright}\small{Imperial-TP-AT-2015-{01}}\end{flushright}

 \maketitle

\flushbottom
\def \De {\Delta} 
\def \ads {AdS$_{5}$\ }
\def \te {\textstyle} \def \iffa {\iffalse} 

\def \ha {{\te {1 \ov 2}}}

 \def  \ba { \begin{align} }
 \def  \ea { \end{align} }

\def \gg   {{\rm g}}
\def \cc    {{\rm c}} 
\def \aa  {{\rm a}}

\def \ep {\epsilon}
 \def \k {\kappa} \def \r {\rho} 

\def \RR {{\rm R}}
\def \OO {{\cal O}} 
\def \edd {\end{document}} 
\def \td {\tilde} 

\def \tO {{\td \OO}}\def \rZ {{\rm Z}}

 \def \Deltat  {{\bar \OO}}

 \def \RR {{\rm R}}
 
 \def \hal {{\te {1 \ov 2}}}
 \def \tot {{\rm tot}}
\def \de {\delta} 
\def \d {{\rm d}}
\def \ed   {\end{document}}
\def \D  {\Delta} 
\newcommand{\hh}{{\textstyle\frac{1}{2}}}

\def \chs {{\rm CHS}} 
\def \mhs {{\rm MHS}} 
\def \eo  {{\rm EO}}

\def \conf {{\rm  CST}}
\def \aa {{\rm a}} 
\def \na {\nabla}

 \def  \ba { \begin{align} }
 \def  \eaa { \end{align} }
\def \no {\nonumber}
\def \va {\varepsilon} \def \m {\mu} \def \n {\nu}   \def \l {\lambda}  \def \k {\kappa} 

\newpage

\section{Introduction}

Higher spin theories containing infinite number of  fields  \ci{Fradkin:1987ks,Vasiliev:2004qz}
may have novel  unexpected   properties  at the quantum  level  related to the fact that one is  required    to sum 
 an infinite number of individual field contributions.  This  summation  requires a particular regularization prescription 
that should be consistent with underlying symmetries   of the theory.  Some examples were   
 discussed  in \ci{Bekaert:2009ud,Gupta:2012he}
 and  especially  in \ci{Giombi:2013yva,Giombi:2013fka,Tseytlin:2013jya,Giombi:2014iua, Giombi:2014yra,
 Beccaria:2014jxa,Beccaria:2014xda}  which we will    elaborate on. 
 
 Our aim will be to study  free partition functions  in flat and conformally flat backgrounds for infinite families 
 of higher spin fields. 
 In addition to the Fronsdal massless  higher spin  (MHS)  fields  with   standard 2-derivative  kinetic terms we 
 will  consider  higher derivative  conformal higher spin (CHS)  fields \ci{Fradkin:1985am}   and 
 also conformal symmetric tensor (CST)  fields with  2-derivative Weyl-invariant actions \ci{Erdmenger:1997wy}. 
 
 We shall start in section 2.1 with a simple but remarkable observation   that the flat-space   partition function 
 of the free  MHS  theory  containing each $s=0,1,2,...$ spin once 
   is trivial, or, equivalently, its   regularized total number of 
 dynamical   degrees of freedom vanishes. This reflects  the presence of a large gauge symmetry, with  the 
 contribution of the determinant of   spin $s$  kinetic operator cancelling against  that of the 
  ghost determinant for  the spin $s+1$ field. This one-loop  $Z=1$  property  generalizes to the 
    $AdS_{\dd}$ vacuum  background of Vasiliev theory   provided  one uses a special  regularization prescription 
 \ci{Giombi:2013fka,Giombi:2014iua}. In particular, we will  find  that for even  dimension $\dd$  there are   special mass sum rules 
 implying the cancellation of not only  logarithmic  \ci{Giombi:2013fka,Giombi:2014iua}   but also  power   UV divergences. 
 
 In section 2.2 we will show     that the $Z=1$ property  holds also in the CHS theory on  flat background: here the
 determinants do not cancel automatically   but the regularized   number of dynamical degrees of freedom still vanishes. 
 As was found in \ci{Giombi:2013yva,Tseytlin:2013jya,Giombi:2014iua}, the regularized sums over $s$ 
 of the conformal anomaly $\aa_s$ and $\cc_s$ coefficients  of  the CHS theory  vanish, implying its  one-loop UV finiteness  4 dimensions  and suggesting that its total 
 one-loop partition  function on a conformally-flat background   should be trivial.  We shall consider the case  of  the 
 CHS theory  on $S^4$ 
 where for each spin $s$   its partition function  should  be   given by the   ratio  \rf{20}  of  MHS  partition functions with alternative boundary conditions    \ci{Giombi:2013yva,Tseytlin:2013jya,Beccaria:2014xda}. This relation will be verified  in Appendix A  following  the dimensional 
 regularization approach  used in the spin 0 case  in  \ci{Diaz:2007an}. Summing over all   spins  implies 
 then that $[Z_\MHS (AdS_5)]_\tot =1$ is directly related to $[Z_\CHS(S^4)]_\tot  =1$.
 
In section 3  we shall study  2-derivative  Weyl-invariant  "higher spin"   actions  for symmetric traceless 
rank $s$ fields in 4 dimensions that have only scalar gauge invariance in conformally-flat background.
This CST   theory is  non-unitary  (though for a different reason than CHS   one) and may be  viewed  as a "maximal depth"  $r=s$ 
representative of a family of conformal higher spin fields   with rank $s-r$ tensor  gauge invariance 
 \ci{Vasiliev:2009ck,Bekaert:2013zya}  (with CHS  case being "minimal depth" case $r=1$).
  In   flat $d=4$ space the  number of dynamical degrees of freedom of rank $s$ CST   field 
 happens to be the same $s(s+1)$ as  of  a CHS  spin $s$ field  and thus the total regularized $Z_{\rm CST}$  is again equal to 1.
 The  associated   $SO(2,4)$   conformal group   representation  is $(3; {s\ov 2}, {s\ov 2})$ that  corresponds 
   to  "maximal-depth"  partially massless \ci{Bekaert:2013zya}   spin $s$ field  in $AdS_5$. 
   We will consider the CST fields   defined on a curved 4d background  and compute  the corresponding partition  function and conformal anomaly $\aa_s$ and $\cc_s$ coefficients comparing them   with the  general expressions for 
   representation $(\Delta; {s\ov 2}, {s\ov 2})$   field  given  in \ci{Beccaria:2014xda}.
    Some details about scalar   gauge invariance in curved space and  CST partition functions on $S^1 \times S^3$ and $S^4$ will be   presented in  Appendices   B,C   and D. 

\section{Summing over spins} 

\subsection{Massless higher spins}

\subsubsection{Flat   space}

Let us consider   the standard  2-derivative   free massless   higher  spin  (MHS)  field in flat  $\d$-dimensional space.  
The  corresponding  partition function can   be written as 
\ba \la{1} 
Z_{{\rm MHS}, s}  = \Big[
\frac{\det\Delta_{s-1\, \perp}}{\det\Delta_{s\,\perp}}
\Big]^{1/2}  = \Big[ \frac{(\det\Delta_{s-1})^{2}}{\det\Delta_{s}\det\Delta_{s-2}}
\Big]^{1/2} 
\ ,    \end{align}
where $\D_s$  is  flat Laplacian  $ -\del^2$  defined on symmetric  rank $s$ traceless tensors, and 
$\Delta_{s\perp}$   is its restriction to transverse fields. We shall assume that $\det\Delta_{k}$     with  $k <0$ is replaced by 1, 
i.e.   $Z_{{\rm MHS}, 0} =(\det \Delta_0)^{-1/2}$. 
Let  us  consider a theory   where  each massless   field  with $s=0,1,2,...$   appears just once. 
This is  field content of   Vasiliev theory  linearized near $AdS_\d$  vacuum  which is dual  to a  large $N$ free 
complex  scalar theory in $d= \d-1$  dimensions. While the expansion of  interacting 
Vasiliev  theory near flat space is singular,
 we may formally view  the free  MHS   partition function  in flat space as a  formal 
   zero curvature limit of its one-loop  counterpart in  $AdS_\d$. 
 Then  one finds that  the total  partition function  is trivial: 
\be \la{2} 
(Z_{{\rm MHS}})_{\tot}= \prod_{s=0}^\infty Z_{{\rm MHS}, s} = \Big[
\frac{1}{\det\Delta_{0}}\Big]^{1/2}  \Big[
\frac{\det\Delta_{0}}{\det\Delta_{1\,\perp}}\Big]^{1/2} \Big[
\frac{\det\Delta_{1\,\perp}}{\det\Delta_{2\, \perp}}\Big]^{1/2} \Big[
\frac{\det\Delta_{2\,\perp}}{\det\Delta_{3\, \perp}}\Big]^{1/2}... =1   \ .  \ee
This remarkable  property   reminds  of a supersymmetric theory   where   the  bosonic contribution to the vacuum partition  function is  cancelled against the fermionic one (implying  the vanishing of the vacuum energy). 
 Here  the cancellation  is   between  
the  contribution of the  physical  spin $s$   field determinant   and the ghost  determinant  for spin $s+1$ field, i.e. 
it reflects a  large gauge  symmetry  of the theory. \foot{This  also suggests  an  analogy with a topological theory. 
Similar examples  are   an antisymmetric  tensor potential of rank $d$ in  $d+1$ dimensions, Chern-Simons theory   and    3d gravity.} 

The cancellation of an infinite  number of factors   in \rf{2}  is  formal  (cf.  1-1+1-1+...=0) as it depends 
on  how one groups terms together:  in general,  an  infinite product requires a regularization  and  depends on 
its  choice.  The choice of regularization should   be consistent with an underlying symmetry  of the theory (in the present  case -- higher spin gauge symmetry).   Let us first consider   the case of $\d=4$.   Observing that each  spin $s>0$ field    has 2 dynamical degrees of freedom  (cf. \rf{1}) 
\be \la{3} 
Z_{{\rm MHS}, s} = (Z_0 )^{\nu_s}  \ , \ \ \ \ \ \   Z_0  = \Big[ \frac{1}{\det\Delta_{0}}\Big]^{1/2} \ , \ \ \ \ \ \ \ 
\nu_s = (s+1)^2 + (s-1)^2 - 2 s^2 = 2  \ ,   \ee
we get 
\be \la{4}
Z_{\tot}=(Z_0 )^{\nu_\tot}\ , \ \ \ \ \ \ \ \ \ \  \qquad   {\nu_\tot} = 1   + \sum_{s=1}^\infty \n_s = 1 +  2   \sum_{s=1}^\infty   1 =0  \ ,  
\ee
where   we used  the standard Riemann  zeta-function as a regularization of the sum over $s$: 
$ \sum_{s=1}^\infty   1= \zeta_R(0) = - {1\ov 2} $. 
Remarkably, in $\d=4$ the use of  the   simple zeta-function 
 regularization is thus equivalent to the formal cancellation of  factors in \rf{2}.\foot{This is  also 
 reminiscent of the  use of the zeta-function  regularization in computing vacuum   energy    in  bosonic  string theory, 
where the use of $\zeta_R(-1) = - {1\ov12} $   leads to the value of the  tachyon mass  ensuring that 
the  vector particle appearing on the first excited level is massless in $D=26$, in agreement with symmetries  of the critical  string   theory.} 
Note that this   flat-space  property $(Z_{{\rm MHS}})_{\tot}=1$ is non-trivial, i.e. it  holds   for a  flat 
torus,  implying,   in particular,  the vanishing of vacuum energy or 
  finite temperature  partition function  on $S^1 \times R^3$. 
  The  partition   function may be non-trivial if one considers an orbifold of  flat space.

For  a   massless   spin $s$  field   in $\d$    flat dimensions   we get 
\ba \la{5}
 \det \Delta_{s} = ( \det \D_0)^{N_s} , \ \qquad   & \det \Delta_{\perp\,s} = (\det \D_0)^{N^\perp_s } \ ,  \\
 N_{s} = \te \binom{s+\d-1}{s}-\binom{s+\d-3}{s-2}\ ,  \quad &N_{s}^{\perp} = N_{s}-N_{s-1}, \ \ \ 
    \nu_s = N_{s}^{\perp} -   N_{s-1}^{\perp}  = \te  2  \big[ s + {1 \ov 2}  (\d-4)\big]  { ( s + \d -5)! \ov s! (\d-4)!} 
   \no \end{align}
   Then $Z_{\tot}=1$   or $ {\nu_\tot} = 0$ is true, e.g., for general  even $\d$   if one uses  the following regularization: 
\be \la{6} 
\nu_\tot  =1 + \sum_{s=1}^\infty \nu_s \ e^{- \ep  [ s + {1 \ov 2}  (\d-4) ]   }\Big|_{\rm fin.}  =0  \ . 
\ee
Here one   performs the sum for  fixed $\ep$, then  takes $\ep\to 0$ and finally  drops  all singular  $ 1 \ov \ep^n$ terms
(in $\d=4$ this  is equivalent to the standard zeta-function prescription).\foot{
An alternative regularization  that gives  vanishing result in any $\d$ is to introduce a cutoff function 
$f(s,\ep)$ (with $f(s,0)=1$)  for each  $\D_{\perp, s}$   factor  in \rf{1} separately  
thus getting  
\be\no   
\nu_\tot  = 1 + \sum_{s=1}^\infty  \Big[   f(s,\ep)\  N_{s}^{\perp} -    f(s-1,\ep) \  N_{s-1}^{\perp} \Big]  =0  \ . 
\ee
This prescription  is the direct analog  of the   cancellation of the determinant factors in \rf{2}. }

\subsubsection{Ricci-flat  space} 

One may   wonder if the $(Z_{{\rm MHS}})_{\tot}=1$ property may generalize to   curved spaces, e.g., Ricci-flat ones. 
As is well known,      massless  higher spin  theories   are not consistent in $R_{\m\n}=0$   background   (do not have flat-space gauge symmetries   surviving) for $s>2$. However,  one may formally   assume that there   exists a consistent  theory of all  higher   spin 
fields where  proper  gauge symmetry is present off-shell  at interacting level.  
Vasiliev theory  does  not   have  $R_{\m\n}=0$ as a 
  classical  vacuum solution 
  since  the limit of vanishing cosmological constant appears to be singular in the 
  interaction terms,  but one may  consider  formally expanding the (hopefully existing) action of Vasiliev  theory near 
an off-shell   $R_{\m\n}=0$    background and computing   the resulting  one-loop partition function.  If $R_{\m\n}=0$  is not a    classical 
solution this partition function will be gauge-dependent, but otherwise   it 
may be  well-defined and  of    interest being a direct  generalization of the flat-space one \rf{1},\rf{2}.
A natural    spin $s$   counterpart of  spin 2 Lichnerowicz operator  on Ricci flat background 
     may be chosen  as in 
\ci{Christensen:1978md} \foot{This is of course a  strong  assumption
(motivated   just by simplicity)   as a   generalization of the Lichnerowitz
operator coming out of a  consistent   higher spin theory formally
expanded near a  Ricci-flat background may contain also    higher
derivative  terms   with    higher powers of   the curvature tensor,
cf. \ci{Zinoviev:2008ck,
Bekaert:2010hw}.   }
\be\la{55}
\Delta_{{\rm L}\, s} = -\nabla^{2}_s+X_s\ ,\qquad  \qquad (X_s\,\varphi)^{\m_{1}\cdots \m_{s}} = -s(s-1)\,R_{\n}{}^{(\m_{1}}
{}_{\lambda}{}^{\m_{2}}\varphi^{\m_{3}\cdots \m_{s})  \n\lambda}.
\ee
Then  one may formally consider the   following   generalization of  the  well-known 
spin $s=1,2 $ partition  functions  on Ricci-flat background to  any spin $s$   \ci{Tseytlin:2013jya} 
\be\la{44}
Z_{{\rm MHS}, s} = \Big[
\frac{(\det\Delta_{{\rm L}\, s-1})^{2}}{\det\Delta_{{\rm L}\,s}\det\Delta_{{\rm L}\,s-2}}
\Big]^{1/2}.
\ee
 Taking the product of \rf{44}   over $s$  we again get  that $(Z_{{\rm MHS}})_{\tot}=1$ as in the  flat space  case  \rf{2}:
\ba \la{2233} 
&(Z_{{\rm MHS}})_{\tot}= \prod_{s=0}^\infty Z_{{\rm MHS}, s}  \\
&\ \ \ \ = \Big[
\frac{1}{\det\Delta_{{\rm L}\, 0}}\Big]^{1/2} \     \Big[ \frac{(\det\Delta_{{\rm L}\, 0})^{2}}{\det\Delta_{{\rm L}\,1 } }
\Big]^{1/2}   \Big[ \frac{(\det\Delta_{{\rm L}\, 1})^{2}}{\det\Delta_{{\rm L}\,2}  \det\Delta_{{\rm L}\, 0}     }
\Big]^{1/2}  \Big[ \frac{(\det\Delta_{{\rm L}\, 2})^{2}}{\det\Delta_{{\rm L}\,3}  \det\Delta_{{\rm L}\, 1}     }
\Big]^{1/2}   ....             =1 
    \ . \no    \end{align}
Here  the cancellation   follows  from the fact that 
  each   spin $s$ operator appears   exactly  twice in both the  numerator and the denominator.

\def \vab {\sigma} 


\subsubsection{Conformally-flat  case:   $AdS_{\d}$} 

The $Z_\tot=1$ property is expected to   hold  also in the  proper vacuum of the  Vasiliev theory  --  $AdS_{\d}$  space, 
and should be true  to all orders in the coupling expansion. 
As was pointed out in \ci{Giombi:2013fka,Giombi:2014iua}, this is the requirement of consistency  of  the vectorial AdS/CFT:
  the boundary theory of  free  $U(N)$ scalar  has log of its  partition function scaling as $N$   which should match 
  the classical   action of the Vasiliev theory in $AdS_{\d}$,  while  the 1-loop  (and all higher-loop)  corrections to 
   $\ln Z_\tot$ of  MHS theory    should  vanish (in a proper regularization). 
     To  demonstrate this even at  the  one-loop   order (free MHS  theory   in $AdS_\d$   background) 
      is, however, much less trivial than in the  flat space background  considered above. 
  Let us   introduce the operator ($ k=0, 1, ...., s-1$)
\be\la{7} 
\D_s (M^2_{s,k}) \equiv - \nabla^2_s + M^2_{s,k} \vab  \ , \qquad  \ \ \qquad 
M^{2}_{s,k} = s-(k-1)(k+\d-2)     \ , \ee
where $\vab=\pm a^{-2} = \pm1$ for unit-radius  $S^\d$  or  euclidean $AdS_{\d}$ space ($\vab=0$ in flat space). 
Let us then  define  the partition function   of a  "partially-massless"  \ci{Deser:1983mm}     spin $s$  field
 (with gauge invariance  with rank $k$  tensor parameter)   \ci{Tseytlin:2013jya}
\be \la{8} 
Z_{s,k} = \Big[
\frac{\det\Delta_{k\, \perp}(M^{2}_{k,s})}{\det\Delta_{s\, \perp}(M^{2}_{s,k})} \Big]^{1/2}  \ . 
\ee
Then for the {\it massless} spin $s$  field (having maximal gauge invariance with  rank  $k=s-1$ parameter) 
on a  homogeneous  conformally flat space we get   the following 
counterpart of \rf{1} 
(see \cite{Gaberdiel:2010ar,Gupta:2012he,Tseytlin:2013jya,Metsaev:2014vda})
\ba
Z_{{\rm MHS}, s}\equiv  Z_{s,s-1} &=\Big[ \frac{\det\Delta_{s-1\, \perp}(M^{2}_{s-1,s})}{\det\Delta_{s\, \perp}(M^{2}_{s,s-1})}
\Big]^{1/2}  \no 
\\ 
& = 
\Big[
\frac{\big(\det\Delta_{s-1}(M^{2}_{s-1,s})\big)^{2}}{\det\Delta_{s}(M^{2}_{s,s-1})
\det\Delta_{s-2}(M^{2}_{s+2,s+1})} \Big]^{1/2} \ , \la{9}
\end{align}
where $s >0$ and for $s=0$     we have 
 $Z_{{\rm MHS}, 0} = [ \det  ( - \nabla^2 + M^2_{0} )]^{-1/2}, \ \  M^2_0=M^2_{0,-1} \vab = 2 (\d-3) \vab$.

 Taking the product of \rf{9} over $s$   
 we conclude that    there is no straightforward   cancellations of determinant  factors  that  happened   in flat space in \rf{2}  or in \rf{2233}: 
 the same-spin operators that appear in the numerator  and denominator of 
$ \prod_{s=0}^\infty Z_{{\rm MHS}, s}$ are different for   non-zero curvature, i.e. $\vab\not=0$: they have  different  
mass-like  terms in \rf{7}. For example, in $\d=4$ case we get 
\be \la{2e} 
(Z_{{\rm MHS}})_{\tot}=
\Big[
\frac{1}{\det\Delta_{0}(2)}\Big]^{1/2}  \Big[
\frac{\det\Delta_{0}(0)}{\det\Delta_{1\,\perp}(3)}\Big]^{1/2} \Big[
\frac{\det\Delta_{1\,\perp}(-3)}{\det\Delta_{2\, \perp}(2)}\Big]^{1/2} \Big[
\frac{\det\Delta_{2\,\perp}(-8)}{\det\Delta_{3\, \perp}(-1)}\Big]^{1/2}...    \ .  \ee
Using   spectral zeta-function regularization one  has, in  a  homogeneous space  background, 
\be  \ln \det  \D_s  = -  \zeta_{\D_s} (0) \ln ( L a)^2   -   \zeta'_{\D_s} (0) \ , \qquad \ \ \ \ \ \ \ \ \ \ \ 
 \zeta_{\D_s} (z)=  \bar  \zeta_{\D_s} (z) \,  V_\d \ .   \la{zeet}   \ee
 Here $L$ is UV cutoff,  $a$ is the curvature radius  and $V_\d$  is the volume. 
 In a  non-compact  space  like $AdS_\d$  where the  volume is formally divergent 
 this relation requires 
 an IR  regularization  (see, e.g.,  \ci{Diaz:2007an} and refs. there). 
 Dropping power IR divergences, regularized   $V_\d$ is then  finite for even $\d$ and  log divergent for odd $\d$
 ($\RR$ is an IR cutoff)
 \be \la{vev}  V_{\d=\rm  even}=  k_1 \ , \qquad \qquad 
 V_{\d=\rm odd}=  k_2  \ln \RR  + k_3 \ . \ee
As  was shown  in \ci{Giombi:2014iua} using the explicit form of higher spin heat kernel   for $AdS_\d$  \ci{Camporesi:1994ga},  
keeping the argument $z$ of spectral $\zeta$-function   non-zero, summing  over spins  and then taking $z\to 0$ 
 one finds that $\bar \zeta_\tot (z) = O(z^2)$, i.e.  that  $ \bar \zeta_\tot (0)=0, \ \ \bar  \zeta'_\tot (0) =0$. 
 Thus   for any $AdS_\d$  with $\d>3$ one has\foot{For odd $\d$   there are no UV   divergences,  i.e. 
 one automatically has $\zeta_{\D_s} (0)=0$, but one is still to show that the finite part  vanishes too, 
 i.e. $\zeta'_\tot (0) =0$.}
 \be\la{99}  (Z_{{\rm MHS}})_{\tot} =    \prod_{s=0}^\infty Z_{{\rm MHS}, s} =  1  \ . \ee
 The same   conclusion should  be  true also   in dimensional regularization used in \ci{Diaz:2007an} (see also Appendix A) 
 where 
 $V_\d= \pi^{\d-1\ov 2} \Gamma( - {\d-1 \ov 2} )$  contains a  pole  for odd $\d\to \d-\va$  (i.e.   ${1 \ov \va }\sim \ln \RR$). Here  
  the   product $\bar  \zeta_{\D_s} (z) \,  V_\d $ in \rf{zeet}   has also a non-trivial   finite part   whose $s$-dependent coefficient need not be the  same as the coefficient of the $1\ov \va$ pole part. However, since $ \sum_s  \bar  \zeta'_{{\rm eff},s} (0)  =0$  property was shown in 
  \ci{Giombi:2014iua}  to be true  for any $\d$,  the multiplication by $s$-independent factor $V_d$   sj=hould not change this   conclusion.


In    general, if  one uses  dimensional (e.g. proper-time)   UV cutoff $L$, then $\ln \det \Delta_s$ in \rf{zeet}  will   contain also power 
  divergences.  The coefficients   of  such $L^{d-2n}$   terms   in $\d$  dimensions   are controlled by 
the Seeley coefficients  or $\zeta_{\D_s} (0)$  functions in $\d'=\d-2n$ dimensions. Since the sum over $s$ of 
the  corresponding combination of $\zeta(0)$ vanishes   in any  $\d>2 $  \ci{Giombi:2014iua}, 
this suggests that  all power divergences should   thus   be absent too,   demonstrating $Z_\tot=1$ in the
 proper-time cutoff regularization. 
This   cancellation of power divergences is again   analogous to what happens in supersymmetric theories. 

For example, in $\d=4$  there  will be $L^4$   divergence   with the  coefficient  of the 
   total number of degrees of freedom (which vanishes   according to \rf{4},\rf{6})
     and also $L^2$ divergence 
   proportional  to $\tr ( {1\ov 6} R - M^2) $ for an operator $-\nabla^2 + M^2$  (here $R=12 \vab a^{-2} $ is the scalar curvature of the background metric). 
   The   coefficient  of the $R$-term  is again  the  total number  of degrees of freedom, 
     while the   contribution of the $M^2$  terms is found to be (we suppress the overall sign factor $\vab$) 
   \be \la{10}
    2 + \sum_{s=1}^\infty (  M^2_{s,s-1}    N^\perp_s  - M^2_{s-1,s}    N^\perp_{s-1} ) 
   =    2 +   4  \sum_{s=1}^\infty  (  1 + 2 s^2)  =0 \ ,  \ee
    where in the last  step we used  the  standard zeta-function   regularization
    equivalent to $e^{-\ep s}$ cutoff.
   In any even $\d$ one finds the same vanishing result using the regularization  
   factor $f(s,\ep) = \exp [ - \ep ( s + {\d-4\ov 2} ) ]$   as in \ci{Giombi:2014iua}.
   
   Explicitly, higher Seeley coefficients for  an operator $-\nabla^2 + M^2$ on a  constant curvature space 
     are given by   $\sum_{k,n} \tr ( R^k M^{2n})$    and thus  are expressed in terms of terms  proportional to 
     $ \tr ( M^{2n})$. 
   Eq.  \rf{10} is interpreted   as    $\tr M^2 =0$;   one  can show   also  the validity of  its higher   power analogs or  mass sum rules 
   $\tr (M^{2n}) =0$,  where 
   \be\la{11}
\text{tr}(M^{2n}) =  [2(\d-3)]^{n}
+\sum_{s=1}^{\infty}
e^{-\epsilon\,(s+\frac{\d-4}{2})}\,
\Big[
M^{2n}_{s,s-1}\,N_{s}+M^{2n}_{s+2,s+1}\,N_{s-2}-2M^{2n}_{s-1,s}\,N_{s-1}
\Big] \ . 
\ee
Here the first term is the contribution from the spin 0 field.\foot{For example, for $\d=4$,
$\text{tr}(M^{4}) =  4-2\,\sum_{s=1}^{\infty}
e^{-\epsilon\,s}\, (s^{2}-4) (5 s^2+1) =-\frac{240}{\epsilon^{5}}+ \frac{76}{\epsilon^{3}}+\frac{8}{\epsilon}+\mc O(\epsilon), $ \ \ 
$\text{tr}(M^{6}) =  8+2\,\sum_{s=1}^{\infty}e^{-\epsilon\,s}\,(2 s^6-23 s^4+67 s^2+8) = 
\frac{2880}{\epsilon^{7}}-\frac{1104}{\epsilon^{5}}+\frac{268}{\epsilon^{3}}+\frac{16}{\epsilon}+\mc O(\epsilon) $. 
In $\d=4$  the exponential regularization is  equivalent to  the  use of the Riemann zeta function  prescription  with 
$\zeta_R ( -2n) =0, \ \zeta_R ( 0) = - \ha$. 
For general $n$  the summand can be written as
$h(s)+h(-s)$, where 
$
h(s) = M^{2n}_{s,s-1}\,N_{s}-M^{2n}_{s-1,s}\,N_{s-1}, $ \  $
h(-s) = M^{2n}_{s+2,s+1}\,N_{s-2}-M^{2n}_{s-1,s}\,N_{s-1}.
$
Then    the only non-vanishing 
contribution in the sum over $s$  is  coming from the  $s^{0}$ term,   giving  $2\,f(0)\,\zeta_R(0) = -2^{n}$,   and this 
 cancels against  the  first term in \rf{11}. 
}

It is natural to   conjecture that the $Z_\tot=1$ property \rf{99}   should   be true not only  for  the 1-loop  partition function, but 
 also  for  the exact   $AdS_\d$    vacuum partition function of the Vasiliev theory
  (i.e. at  any quantum  loop order  in the   $ 1/N$ coupling expansion). 
  As already mentioned,   this  the requirement  of the vectorial AdS/CFT duality:  
 the   logarithm   of  the partition function  of the dual  free $U(N)$   scalar theory  has only the   order $N$  term 
 (that should match  the   vacuum value of the classical  action of the Vasiliev theory). 
 This  further   strengthens the analogy  with a supersymmetric or topological   quantum field theory. 
 
 Finally, let us note that the property \rf{99}    need  not apply to  quotients  of the  $AdS_\d$ space -- for example, 
the MHS partition function on thermal quotient  of $AdS_\d$  is non-trivial (see \ci{Giombi:2014yra}  and refs. there).


\subsection{Conformal higher spins}
\subsubsection{Flat   space}

Let us   now   consider  the  free   partition function for conformal   higher spin  (CHS)   theory 
\ci{Fradkin:1985am,Tseytlin:2013jya}. The   flat space action  for a  free  CHS   field  in $d$ dimensions 
   is  $\int  d^d x \ \vp_s  P_s \del^{2s + d -4 } \vp_s$, where  $P_s$ is projector 
to transverse   traceless   totally symmetric  rank $s$ field.   Here $s=0$   is a non-dynamical scalar, 
$s=1$ is the Maxwell   vector, $s=2$ is the Weyl graviton, etc. 
 The corresponding   partition function in $d=4$    is  \ci{Tseytlin:2013jya}
\be  Z_{{\rm CHS}, s} =
\Big[
\frac{(\det\Delta_{s-1})^{s+1} }{(\det \Delta_{s})^s}
\Big]^{1/2}
=
\prod_{k=0}^{s-1}  \   \Big[
\frac{\det\Delta_{k\, \perp}}{\det\Delta_{s\, \perp}}
\Big]^{1/2}, \la{12}
\ee
where as in \rf{1}  the operator $ \Delta_{s}= - \del^2$  is  defined   on symmetric traceless  tensors.  

CHS   fields having dimension $2-s$ are  sources or "shadow  fields" for  spin $s$  conserved  bilinear currents $J_s(\p)$ 
 built out of a free  $U(N)$  scalar field;  they are also boundary   values for the  corresponding dual  MHS theory in $AdS_{d+1}$. 
An interacting  CHS theory may be  defined as an  induced one \ci{Tseytlin:2002gz,Segal:2002gd,Bekaert:2010ky}, 
 obtained by integrating out  $\phi$ in  the  path integral defined by  the action  $\int d^4 x \big[\del \p^* \del \p   +  \sum_s  J_s(\p) \vp_s\big]$. 
  The  resulting     interacting  CHS theory   contains all fields   with spins $s=0,1,2,...$. 
  The corresponding  free   partition function  in flat background   is given by  
\be \la{13} 
(Z_{{\rm CHS}})_{\tot}= \prod_{s=1}^\infty Z_{{\rm CHS}, s} =    \Big[ \frac{ \det\Delta_{0} }{\det \Delta_{1}} \Big]^{1/2}      
    \Big[ \frac{(\det\Delta_{1})^{3} }{(\det \Delta_{2})^2} \Big]^{1/2}   
      \Big[ \frac{(\det\Delta_{2})^{4} }{(\det \Delta_{3})^3} \Big]^{1/2}  ...     \ .    \ee
Formally cancelling similar   factors  in  the numerator and  the denominator  of \rf{13} 
   leads, in contrast to \rf{2}, to a  non-trivial result 
\be \la{14} 
(Z_{{\rm CHS}})_{\tot} \ \  \to\ \  (Z_{{\rm CHS}})'_{\tot}= \prod_{s=0}^\infty  \det\Delta_{s}  \ . \ee 
This rearrangement of an infinite product  in \rf{13} effectively  corresponds to its  particular regularization. 

Alternatively,  we may  use that each $ Z_{{\rm CHS}, s}$   factor  \rf{12}   may be written as in \rf{3},   i.e. as 
 $(Z_0)^{\nu_s} = [\det \Delta_{0}]^{- \n_s/2}$
where  here $\n_s = s(s+1)$   is the  number of dynamical degrees of freedom of a CHS field in $d=4$. Then 
\be \la{15} 
(Z_{{\rm CHS}})_{\tot}= \prod_{s=0}^\infty   (Z_0)^{\nu_s}  =( Z_0)^{\n_\tot} \ , \ \ \qquad  \ \ \   \n_\tot =\sum_{s=0}^\infty \n_s \ , \ \ \ \ \ \ \ 
 \n_s = s(s+1)
 \ . \ee
The  total number of  CHS degrees  of freedom vanishes   if one uses 
the   regularization      suggested   by the relation to the MHS theory in 
  $AdS_{\d}$   with $\d=d+1$  
(in which  also the total conformal anomaly  vanishes \ci{Giombi:2013yva,Tseytlin:2013jya,Giombi:2014iua}).
 Indeed, doing the sum in \rf{15} with the  $\exp [ -\ep (s + { d-3 \ov 2})]$ cutoff  as in \rf{6}   we get  in  the $d=4$ case  
 \be \la{16}
 \n_\tot =\sum_{s=0}^\infty s(s+1) \ e^{ -\ep (s + { 1 \ov 2})} \Big|_{\rm fin.} = 0 \ , \ \ \ \ \qquad  
 {\rm i.e.} \qquad  \ \ \ \ \ \ \  (Z_{{\rm CHS}})_{\tot}=1 \ , 
  \ee 
   as in  the MHS case in \rf{2},\rf{6}.
   This   conclusion  generalises to the case of the CHS theory in 
 $d$ even dimensions where 
 the partition function is \ci{Tseytlin:2013jya} \ \
 \ba \la{166}
& Z_{{\rm CHS}, s}= \Big[\big(\frac{1}{\det\Delta_{s\perp}}\big)^{d-4\ov 2}\,\prod_{k=0}^{s-1}  \
\frac{\det\Delta_{k\perp}}{\det\Delta_{s\perp}}
\Big]^{1/2}= (Z_0)^{\nu_s} \ ,  \\
& \la{161}
\ \ \ \ \ \ \ \ \  \nu_s = \frac{(d-3) (2 s+ d-4) (2 s+d-2) (s + d-4)!}{2 (d-2)! \ s!}  \ . \end{align}
Then $
 \n_\tot =\sum_{s=0}^\infty \n_s  \ e^{ -\ep (s + { d-3  \ov 2})} \Big|_{\rm fin. } = 0$
 and thus $(Z_{{\rm CHS}})_{\tot}=1$. 
 Notice that  here $d$  is the  dimension of the boundary where CHS theory is defined; the  related MHS theory 
 defined in $\d=d+1$   has equivalent regularization  used, e.g.,  in  \rf{6}. 
This regularization should be  the  one that is consistent with the symmetries of the CHS theory. 

 At the same time,  if we start  with the rearranged product \rf{14}  that  can be written 
 as 
 \be 
 (Z_{{\rm CHS}})'_{\tot}= ( Z_0)^{-2 N_\tot} \ , \ \ \ \ \ \   \ \ \ \    N_\tot = \sum_{s=0}^\infty   N_s \ , \ \ \ \ \ \ \ \ \ \ \ 
 N_s = (s+1)^2  \ , \la{17}
  \ee 
we   find that  $N_\tot = { 1 \ov 24}$    if one uses the same regularization as in \rf{16}.\foot{ $N_\tot$ does   vanish
   in a different regularization:  with the cutoff factor being 
$e^{ -\ep (s + { 1 })}$.}  This   illustrates an  ambiguity     associated with 
  formal rearrangements of an infinite product: the result depends on a regularization.

\subsubsection{Ricci-flat  space} 

In contrast to the   2-derivative massless higher spin theory
  the  CHS  theory is expected to admit  a Ricci-flat (or, more generally, Bach) 
background as its classical solution  and each  CHS   field  should  have   proper gauge invariance
 in   such    background.  Thus  the  free CHS partition function in  $R_{mn}=0$ should be well-defined
 (gauge-independent).
  If one  makes a bold  conjecture (known to be true  for $s=1,2$)\foot{This  conjecture 
  is likely to be wrong for $s = 3$ \ci{Nutma:2014pua} 
   unless the  background curvature  tensor is subject to   further constraints, but   may be true as far as one is allowed to ignore
   terms with covariant derivatives of the curvature (which,  for example,  do not contribute to nontrivial part of the 
   conformal anomaly in $d=4$).}
   that  the $2s$ derivative covariant CHS  operator can be   factorized into a product of standard 2-derivative   spin $s$ 
 Lichnerowicz operators \rf{55}  \ci{Tseytlin:2013jya}:
 \be\la{18}
Z_{{\rm CHS}, s} =
\Big[
\frac{(\det\Delta_{{\rm L}\,s-1})^{s+1}}{(\det\Delta_{{\rm L}\,s})^{s}}\Big]^{1/2}\ .
\ee
The same  rearrangement of the   infinite product  as in \rf{13}  then gives  the  following
Ricci-flat space generalization  of \rf{14} 
 \be \la{19} 
(Z_{{\rm CHS}})_{\tot}  = \prod_{s=1}^\infty Z_{{\rm CHS}, s}  \ \    \to \ \    (Z_{{\rm CHS}})'_{\tot}= \prod_{s=0}^\infty  \det\Delta_{{\rm L}\,s }  \ . \ee 
From \rf{18}   one finds, in particular,  the following  expression for the $\beta_1= \cc-\aa$   conformal anomaly coefficient  (coefficient   of   the $R^*R^*$  term in trace anomaly on Ricci-flat background)
\ci{Tseytlin:2013jya}:
\be 
 \cc_s-\aa_s=  \te  { 1 \ov 720} \nu_s ( 4 - 45 \nu_s   + 15 \nu_s^2)   \ , \qquad  \ \ \ \ \ \   \nu_s = s (s+1)    \ . \la{199} 
\ee
Then $\sum_{s=1}^\infty (  \cc_s-\aa_s)=0$ in the same regularization \ci{Giombi:2014iua} as in \rf{16},

\subsubsection{Conformally-flat  space: $S^4$} 

As was   shown in \ci{Giombi:2013yva,Tseytlin:2013jya,Giombi:2014iua}, the  sum of  the conformal anomaly
 $\aa_s$-coefficients over all $s$   vanishes   (this is true, in particular, in the same regularization as used in \rf{16}).
One might then  expect   that  the total   partition function  of the CHS theory  on a  conformally-flat space should be  
 simply related  to the   one  in flat space.   
 For example, 
 $Z_\tot(S^4)$   with the product over $s$ computed with the regularization in \rf{16}
  may  again  be  equal to 1.  
  
 This   would   be    consistent with  the  relation between the  spin $s$ MHS partition function  in $AdS_5$ 
   and the  corresponding CHS  partition function on $S^4$ \ci{Giombi:2013yva,Tseytlin:2013jya,Beccaria:2014xda} (see also \ci{Barvinsky:2005ms,Barvinsky:2014kta})
\be\la{20}
Z_{\CHS,s}(S^{4}) = \frac{Z_{\MHS,s }^{-}(\text{AdS}_{5})}{Z_{\MHS,s }^{+}(\text{AdS}_{5})} \ . 
\ee
Here $Z_{\MHS, s }^{+}$ is the  massless  higher spin  partition  function with the standard (Dirichlet) b.c. (as  assumed  
 in \rf{9})
while $Z_{\MHS,s }^{-}$ is its  alternative (Neumann) b.c.   counterpart. 
The relation \rf{20}  should   be true for any $s=0,1,2,...$ and   should thus also   apply also to  the total products  over $s$. 
The computation of  the spectral  zeta-function in  \ci{Giombi:2014iua}  (in which the infinite $AdS_5$ volume  was assumed to factorize uniformly)
implies \rf{99}, i.e. $(Z_{\rm MHS}^{+}(\text{AdS}_{5}))_\tot=1$   
and   also  $(Z_{\rm MHS}^{-}(\text{AdS}_{5}))_\tot=1$.\foot{The condition  $\zeta_{\rm MHS}(0)=0$  is automatic 
in the  $AdS_5$  case, while 
 $\zeta'_{\rm MHS}(0)=0$  is valid for both  choices of the boundary conditions.}
Equivalently, these properties hold in the  same regularization  of the sum over spins as used in   \rf{16}.
Assuming the validity  of \rf{20}   one  should  then  expect to  find  that  in this   summation  
prescription  
\be\la{21}
\big(Z_{\rm CHS}\big)_\tot (S^{4})=\prod_{s=1}^\infty Z_{\CHS,s}(S^{4}) = 1   \ . \ee 
The verification of \rf{20}    by directly computing    the determinants  appearing  on the  both sides of the equality 
turns out to be non-trivial.
The expression
for $Z_{\CHS,s}(S^{4})$   depends on a choice of UV  regularization, while $Z_{\MHS,s }^{\pm}(\text{AdS}_{5})$
depends on a choice of  IR regularization, and these  regularizations 
 should be properly  coordinated  for \rf{20} to hold. 
The question  of how to do this was  previously  addressed only in the  spin  0  analog of the relation \rf{20}  in \ci{Diaz:2007an,Diaz:2008hy} 
where $Z_{\CHS,s}$   is replaced by the  partition function of the  order  $2r$ GJMS operator \ci{Graham:1992}   and 
$Z_{\MHS,s} $ -- by the  $AdS_5$   partition function of  the  massive scalar     with $m^2 = \Delta (\Delta-4) = r^2 -4$. 
We shall  discuss  this case in  detail  Appendix A.1 below. 
For  $s>0$ the relation  \rf{20}  was verified  for the leading  singular (logarithmically divergent)  parts only:
 the spin-dependent coefficient  \ci{Giombi:2013yva}
  of the  IR  divergent $\ln \RR$  term (cf. \rf{vev}) 
    in $\ln  \frac{Z_{\MHS,s }^{-}(\text{AdS}_{5})}{Z_{\MHS,s }^{+}(\text{AdS}_{5})} $  indeed 
matches the  (conformal anomaly $\aa_s$) coefficient  \ci{Tseytlin:2013jya} of the UV divergent $\ln  \Lambda$
  term in $\ln Z_{\CHS,s}(S^{4}) $.\foot{Here $\Lambda $ stands for UV cutoff  in the $d=4$  CHS theory
where determinants   have  similar expression as in \rf{zeet}.}
 
If one  uses   an    IR regularization in $AdS_5$   in which the volume universally factorizes   as in \rf{zeet} 
then the  coefficients of the  $\ln \RR$ and finite parts  in  the MHS   side of \rf{20}   appear to  have  the same   spin dependence. 
This is  certainly not so a priori  for the coefficients of the  UV  divergent and finite parts on the  CHS side of \rf{20}.\foot{One may  of course  formally   absorb the finite  part  into a redefinition of UV   cutoff $\Lambda\to \td \Lambda$    
but that will make $\td \Lambda$ spin-dependent,  precluding its identification  with $\RR$ on the $AdS_5$  
side and  also complicating  the issue   of summation over $s$.} 
As was  pointed out in the   $s=0$ case in \ci{Diaz:2007an}, to be able to  systematically  match the finite parts 
of the partition functions on $S^d$ and $AdS_{d+1}$ one may use dimensional regularization, i.e. $d\to d-\va$. 
Then for even $d$  the regularized   $AdS_{d+1}$   volume      $\pi^{d\ov 2} \Gamma( - {d\ov 2})$   will have $1\ov \va$ pole  term 
 that may hit an order $\va$ term in $\bar \zeta_{\Delta_s}$  in  \rf{zeet} to  produce
  a non-trivial  finite  contribution  that  may  match the finite term  present  in the $S^d$ partition function. 
Generalizing the  dimensional   regularization approach  of \ci{Diaz:2007an} to spin $s > 0$   case on $AdS_5$ side,  
in Appendix A.2  we shall  demonstrate the matching of the  most non-trivial transcendental 
finite parts of the  CHS and MHS sides of \rf{20}.
We shall  then  provide  a   check of  the validity of  the $(Z_\CHS)_\tot=1$ property  \rf{21} 
in this   regularization  consistent   with  \rf{20}    using the same summation over spins  prescription as in 
\rf{16}. 

 Explicitly, the   CHS   partition function  on $S^4$    is given by the following  generalization of \rf{12} 
(cf. \rf{7},\rf{8},\rf{9}) \ci{Tseytlin:2013jya}
\be\la{22}
Z_{{\rm CHS}, s}(S^{4}) = \prod_{k=0}^{s-1} Z_{s, k}\ , \ \ \ 
Z_{s,k} = \Big[ \frac{\det\Delta_{k\, \perp}(M^{2}_{k,s})}{\det\Delta_{s\, \perp}(M^{2}_{s,k})}
\Big]^{1/2}\ , \ \ \  M^{2}_{s,k}\Big|_{d=4} = s - (k-1) (k+2)\ . 
\ee
For the corresponding  conformal  anomaly $\aa$-coefficient one finds 
\be 
\aa_s =\te   { 1 \ov 720} \nu_s ( 3\nu_s   + 14 \nu_s^2)   \ , \qquad  \qquad  \ \ \ \ \   \nu_s = s (s+1)    \ , \la{229} \ee
and then $\sum_{s=1}^\infty \aa_s   =0$ in the   same regularization as  in \rf{16}.\foot{Together   with the vanishing 
of the sum of $\cc_s-\aa_s$ in \rf{199}  this implies the one-loop  UV finiteness of the CHS  theory on a curved 4d background.}
Using the known spectra   of  the  second-order Laplace operators  on $S^4$  one may also compute  explicitly  the finite part of the 
determinants in \rf{22}\foot{We shall  ignore the (rational)   contribution of the multiplicative anomaly, i.e. 
the ratio of $\det (O_1 ... O_n)$ to $\det O_1 ... \det O_n$\ 
\ci{Dowker:2015rta}. The role   of this anomaly in CHS theory is unclear: 
it  does not appear in the formulation based on  introducing auxiliary fields  to write the CHS action in terms of second  derivative operators only \ci{Metsaev:2007rw}.  It is possible   that ignoring this anomaly   is  required  for consistency with the ratio of the MHS partition functions 
in  \rf{20}.}    and   match  the   CHS  expression    with  the   MHS one in \rf{20}   (see  Appendix A.2).

An analog   of  \rf{21} may be expected  to 
   be true  for any conformally-flat space, e.g., for $\mathbb  R \times S^3$ (indeed,  the vanishing of the   CHS 
Casimir energy on $S^3$  was demonstrated  in \ci{Beccaria:2014jxa}). 
At the  same time,   this  need not apply to orbifolds of conformally-flat space:  for example, 
the  finite temperature  CHS partition function on $S^1_\b \times S^3$ is non-trivial   \ci{Beccaria:2014jxa}
(and is  equal  to   the ratio \rf{20}  of  the  non-trivial MHS partition  functions on thermal quotient of $AdS_5$).\foot{While 
$S^1_\b \times S^3$  is locally  conformally flat (the Weyl tensor vanishes)   this is not true globally: the periodicity of $S^1$ coordinate is important. 
Indeed,  to show that $\mathbb{R}\times S^{3}$ is conformal to flat space one   uses  that 
$dx^2 + dS^{3} = y^ {-2} ( dy^2 + y^2 dS^{3}) =   y^ {-2}  \, dy_n dy_n $.  
 If  $x$ is an angle  variable  we cannot  set it equal  to $\ln y$ globally.}


\section{2-derivative  conformal  symmetric tensor   theory } 

Let us now   consider  another     family   of    "higher-spin" (symmetric traceless  tensor) actions 
  \ci{Erdmenger:1997wy}   in 4  dimensions 
 which, like the CHS  ones   are    covariant under   Weyl rescaling of 
 the background metric  but like  the MHS ones  (which are not conformal for $s>1$) are only second 
 order in derivatives. 
 We shall refer to this   theory  as    "conformal   symmetric tensor" (CST) theory. 
  
  These CST fields   are non-unitary  lacking  full higher spin gauge invariance:
  even in (conformally) flat space limit  they have only scalar gauge invariance.
  In fact, they  can be interpreted  as "maximal depth"  (minimal   gauge invariance) 
  representatives  of  a   family of conformal fields
  (called   "FT fields"  in \cite{Bekaert:2013zya})  containing   CHS   fields as maximal gauge invariance members. 
  Like  CHS fields are associated  with massless  fields in $AdS_5$,  CST  fields  in 4d correspond to 
    "maximal depth"  partially massless  fields  in $AdS_5$
     \cite{Bekaert:2013zya}.\foot{As discussed in  \cite{Bekaert:2013zya}, one  may  consider a generalize triple   of  families of fields 
     (here  $d=4$):
     
      (i) 
     starting with higher-derivative   conformal   scalars  in $R^4$   (higher-order singletons)  
with the action $\int d^4 x \ \phi^*_i  ( \del^2)^{\ell} \phi_i $  one may construct  dimension $\Delta= 3 + s-t$
partially conserved currents $J_s$ of spin $s$ and  depth $t$ (with $1 \leq t\leq s$),   
$ \del_{m_1} ...\del_{m_t} J^{m_1...m_s} =0$;  
 the corresponding 
Verma module  $V(\Delta, s) = V(3 + s-t,s)$ is  reducible for $ t \leq s$ 
 and the 
irreducible module   is $D(3+s-t,s) = V(3 + s-t,s)/V(3 + s,s-t) $  \ci{Dolan:2001ih,Shaynkman:2004vu};

(ii) the  sources for these currents or  the corresponding   "shadow" fields are 
primary conformal   fields  of depth $t$  which  are   
totally symmetric traceless  tensors $\vp_{m_1...m_s}$ of dimension    $\Delta= 1+ t-s$
with the action   $\int  d^4 x \ \vp_s  (\del^{2})^{1 + s-t}\vp_s  $;

(iii) these are  naturally associated   to 
 partially massless  fields in $AdS_5$  with the action  $\int d^5 x \p_s (\nabla^2 +...) \p_s $  
having  gauge  invariance 
$ \delta \phi_{\mu_1 ..\mu_s} = \nabla_{\mu_1} ...\nabla_{\mu_t} \epsilon_{\mu_{t+1} ... \mu_{s} }+  (g_{\m\n}-{\rm terms}) $;
the fields $\p_s$   are 
dual to $J_{m_1...m_s}$ or   have $\vp_{m_1...m_s}$ as   boundary values. 

The minimal depth  $t=1$ case   corresponds to    conserved currents, $\vp_s$ as  CHS  fields in $d=4$  and  
$\p_s$ as  MHS   fields in $AdS_5$. 
In  the 
 maximal  depth case  $t=s$  which we will  be considering below $\vp_s$  correspond to the    CST  fields 
  while $\p_s$ -- to partially massless   fields of  maximal depth. 

The 
 field content of  the $AdS_5$   theory dual to   singlet sector of $U(N)$  $(\del^2)^\ell$ scalar theory  follows from 
 the 
 generalized Flato-Fronsdal theorem  \cite{Bekaert:2013zya}:
$ D( 2 -\ell,0) \otimes D( 2 -\ell,0) = \oplus^\infty_{s=0}  \oplus^\ell_{k=1} 
D(4 + s - 2k, s) $, 
where the  sum goes  over 
  partially massless  fields of  different  odd depths $t= 1, ..., 2\ell -1$.
   Thus in contrast to  the   MHS fields  in  the 
    minimal depth case  the maximal  depth $t=s$  fields $\p_s$  do not form   a "closed"  subset, 
    i.e.  one is to group together   fields of different   depth to get the  dual $AdS_5$ theory.
}

  Below we shall first   introduce the corresponding bilinear action in curved background, 
  then    discuss in detail   the    spin  2 case, and   finally     consider the   partition  function 
  for   the  general   spin $s$ case.

\subsection{Classical  Lagrangian}

Given  a  rank $s$   totally symmetric  traceless   tensor  $\vp_s$ there  are  several  options to construct an action invariant 
under   Weyl  rescalings of the metric   combined with a rescaling of the  field $\vp_s$.
If  the action  contains $2n$   derivatives, i.e.  starts  with 
\be \int d^d x\  \sqrt g\ \vp_s  (\nabla^2)^n  \vp_s + ...  \ , \qquad n=1,2,.... \ , \la{nnn}    \ee
then   one should require the invariance under  ($\Omega=\Omega(x)$) 
\be \la{1.1} 
g_{\mu\nu}' = \Omega^{2}\,g_{\mu\nu}\ , \qquad 
\varphi_{\mu_{1}\dots  \mu_{s}}' = \Omega^{ \g  }\,\varphi_{\mu_{1} \dots  \mu_{s}} \ , \ \ \ \ \    \g = s + n - \ha {d } \ . 
\ee
For example, in the CHS case $n=  s + \ha (d-4) $   and $\g= 2s   -2 $. 
 The  family of conformal   operators with $n=1,2, ...$ 
   generalises the  scalar GJMS \ci{Graham:1992}    one  to the  $s> 0$  case. 
In the case of  2-derivative kinetic term ($n=1$)
corresponding to CST   theory 
  one has $\g=s - \ha (d-2)$.  
    The  larger the  $n$, the more gauge symmetries  consistent with locality of the action   one  may 
    expect  to be  realisable (at least, in the flat space  limit).  Indeed, in the CHS case   one 
   has  maximal  gauge symmetry with  rank $s-1$ tensor parameter   while   in the
    $n=1$ case  there will be only  $\delta \vp_s = \del^s \sigma$   gauge symmetry with scalar gauge parameter.  
   
    Having less than   maximal  gauge symmetry,  
     one    will be  unable to   eliminate   all the time-like field 
    components   (contributing  to the action with negative sign)   by a choice of a "unitary" gauge ($n > 1$ actions  will be non-unitary also due to   higher derivative kinetic term). 
     While the standard 2-derivative   
     Fronsdal  massless (maximally gauge-invariant ) higher spin 
    action  is unitary  but not conformally invariant, the  $n=1$  conformally invariant CST action will   not be unitary   due to 
    insufficient  gauge symmetry. 
  
 The  $n=1$  action  invariant under   \rf{1.1}  was  found  in 
    \cite{Erdmenger:1997wy} (cf. also \ci{Iorio:1996ad} for $d=4$):
      it is   given by      
       $S_{s}  = -\frac{1}{s!}\,\int d^{d}x\,\sqrt{g}\,\mathscr L_{s}$  where
\be
\label{1.7}
\begin{split}
\mathscr L_{s}(d)  &= \te 
\na^{\lambda}\varphi^{\mu_{1}\cdots\mu_{s}}\na_{\lambda}\varphi_{\mu_{1}\cdots\mu_{s}}
-\frac{4s}{2s + d-2}\,\na_{\rho}\varphi^{\mu_{1}\cdots\mu_{s-1}\rho}\na_{\lambda}
\varphi_{\mu_{1}\cdots\mu_{s-1}}{}^{\lambda}\\
&\ \ \te  +\frac{2s}{d-2} R_{\rho\lambda}\varphi^{\mu_{1}\cdots\mu_{s-1}\rho}\varphi_{\mu_{1}\cdots\mu_{s-1}}{}^{\lambda}
-\frac{4s - d^{2}+4d-4}{4(d-1)(d-2)}\,R\,\varphi^{\mu_{1}\cdots\mu_{s}}
\varphi_{\mu_{1}\cdots\mu_{s}}\\
&\ \ +\omega\,C_{\alpha\beta\rho\lambda}\,\varphi^{\mu_{1}\cdots\mu_{s-2}\alpha\rho}\,
\varphi_{\mu_{1}\cdots\mu_{s-2}}{}^{\beta\lambda} \ , \qquad \mathscr L_{\conf,s}\equiv  \mathscr L_{s}(d=4) \ . 
\end{split}
\ee
Here $C$  is the  Weyl tensor and $\om$   is an arbitrary constant. 

In what follows   we shall consider the $d=4$  case. 
In  this case  the  flat space   limit  of \rf{1.7}   is invariant under the scalar gauge   transformations\footnote{
In general dimension $d$, the scalar gauge invariance is  present in the case when  the CST kinetic operator 
contains  $ 2 + (d-4) $  derivatives.  Such conformal field  (see   \cite{Shaynkman:2004vu,Vasiliev:2009ck})
 is   the  maximal depth FT field  in \cite{Bekaert:2013zya}.
} 
\be \la{117} 
\te  \mathscr   L_{\conf,s} = 
\del^{\lambda}\varphi^{\mu_{1}\cdots\mu_{s}}\del_{\lambda}\varphi_{\mu_{1}\cdots\mu_{s}}
-\frac{2s}{s +1} (\del^{\lambda}\varphi_{\mu_{1}\cdots\mu_{s-1}\lambda})^2  \ , \ \ \ \ \ \
\delta \varphi_{\mu_{1}\cdots\mu_{s}} = \del_{\m_1} ... \del_{\m_s} \sigma  - {\rm traces} 
\ . \ee 
The field $\vp_s$ has  canonical  dimension 1  and thus  corresponds to a (non-unitary) 
representation  $(1; {s\ov 2}, {s\ov 2}) $ of $SO(2,4)$.\foot{We use the notation 
$(\Delta; j_{1}, j_{2})$ where   $\Delta$ is the scaling 
dimension or conformal weight  and  $(j_{1}, j_{2})$ are the  $SU(2)$  weights of  $SO(3,1)$.
}
This  representation is unitary ($\Delta \geq 2+j_{1}+j_{2}$) only for $s=0,1$. 
 The gauge  parameter   $\sigma$  corresponds to the $d=4$ conformal group  representation $(1-s; 0,0)$, 
 so that \rf{117} describes a "short" representation\foot{This  case should   correspond to a particular 
 degenerate module of the conformal group  \ci{Shaynkman:2004vu,Vasiliev:2009ck,Bekaert:2013zya}.} 
 \be 
 \la{777} \te 
 [1; {s\ov 2}, {s\ov 2}]= (1; {s\ov 2}, {s\ov 2}) - (1-s; 0,0)  \ . \ee 
In $d=4$  and  $s=1$   \rf{1.7}   gives  the standard  Maxwell    Lagrangian.
Let us discuss in detail the spin 2 case.

\subsection{Rank  2 case}

   In  $d=4$ the $s=2$  case  of  \rf{1.7} is 
\be
\label{1.10}
\begin{split}\te 
\mathscr  L_{\conf,2} = \na^{\lambda}\varphi^{\mu\nu}\,\na_{\lambda}\varphi_{\mu\nu}
&\te -\frac{4}{3}\, (\na_{\m}\varphi^{\mu\n} )^2
+2\,R_{\rho\lambda}\,\varphi^{\mu\rho}\,\varphi_{\mu}^{\ \lambda} \\
&\te  -\frac{1}{6}\,R\,\varphi^{\mu\nu}\,\varphi_{\mu\nu}
+\omega\,C_{\mu\nu\rho\lambda}\,\varphi^{\mu\rho}\,\varphi^{\nu\lambda}.
\end{split}
\ee
For comparison,  the  quadratic term  in the expansion of the Einstein action in generic background  is
(here  $g_{\m\n} \to  g_{\m\n}  + h_{\m\n}$,\  \  $\varphi_{\mu\nu} = h_{\mu\nu}-\frac{1}{4}g_{\mu\nu}h$,\  \  $h=h_{\mu}^{\mu}$)
\be
\begin{split}\la{110}
\mathscr L_{\rm E} =  \na^{\lambda}\varphi^{\mu\nu}\,\na_{\lambda}\varphi_{\mu\nu}
&\te - 2\,\Big[ \na_{\m} (\varphi^{\mu\n}-\frac{1}{2}\,g^{\mu\n}\,h )
\Big] ^2  +   { 1 \ov 4}     h \na^ 2  h  \\ 
&\te 
+\frac{5}{3}\,R\,\varphi^{\mu\nu}\,\varphi_{\mu\nu}-2\,C_{\mu\alpha\nu\beta}\,\varphi^{\mu\nu}\,\varphi^{\alpha\beta}\\
\end{split}
\ee
The action  for \rf{110}  has the standard (vector-parameter) gauge invariance  on  a $R_{\m\n}=0$   background, 
while this is not the case for \rf{1.10}. 

We shall refer to \rf{1.10}   as  conformal   spin 2   Lagrangian. 
Let   us   mention   some earlier work related to it. 
As is well known,   in contrast to massless   spin 0 and spin 1   fields, the  spin 2 
 Einstein   graviton    does  not  represent a  conformal  theory  in flat space
 \cite{Grishchuk:1980hb}.\foot{This   follows, e.g., from  the   absence of Weyl invariance of the Einstein theory. 
 In particular,  the conformal weights do not match:
 according to \rf{1.1},   the conformal spin 2  field $\varphi_{\mu\nu}$ is to
have Weyl weight 1 while  the metric itself transforms with weight 2.
For all  massless   spins  $s \geq 2$  one has scale   invariance but no special conformal invariance. 
%
}
The  $SO(2,4)$   conformal group  representation  for the symmetric traceless tensor  $\varphi_{\mu\nu}$ 
is  $(\Delta; j_{1}, j_{2}) = (1; 1, 1)$.
The corresponding    free conformally invariant equations of motion   
\cite{Drew:1980yk} (see also \cite{Barut:1982nj,Anselmi:1999bu})
\be
\label{1.2}
\te \del^2 
\varphi_{\mu\nu}-\frac{4}{3}\Big(\partial_{\alpha}\partial_{(\m} \,\varphi^{\alpha}_{\nu)}-
\frac{1}{4}\,g_{\mu\nu}\,\partial_{\alpha}\partial_{\beta}\,\varphi^{\alpha\beta}\Big)
= 0 
\ee
 follow from the flat-space   Lagrangian \rf{1.10}  or \rf{117}      with 
  scalar gauge invariance 
\be
\label{1.3} \te   L_{\conf,2}=  \del^{\lambda}\varphi^{\mu\nu}\,\del_{\lambda}\varphi_{\mu\nu}
-\frac{4}{3}\, (\del_{\m}\varphi^{\mu\n} )^2\ , \qquad \qquad 
\delta \varphi_{\mu\nu} = \big(\partial_{\mu}\partial_{\nu}-\frac{1}{4}\,g_{\mu\nu}\,\partial^{2}\big)\,\sigma.
\ee
$\vp_{\m\n}$   corresponds to    $(1; 1,1)$ conformal  group representation  which is non-unitary
 \cite{Fang:1982ks}.
  As already mentioned,   non-unitarity  is implied   also 
by  the absence of the maximal (vector-parameter)  
spin 2 gauge invariance:  
in fact, \rf{1.2} describes a combination  of spin 2  and two spin 1  
massless on-shell fields   with   total    $(9-1) -2 =6$  physical     degrees of  freedom. 
Splitting 
\be 
\la{1.5}\te   \varphi_{\mu\nu} = \varphi_{\mu\nu}^{\perp}+\partial_{(\mu}\,V^\perp_{\nu)}  + 
\big(\partial_{\mu}\partial_{\nu}-\frac{1}{4}\,g_{\mu\nu}\,\partial^{2}\big)\,\sigma
\ee
 and accounting for the Jacobian of this transformation and  the  decoupling of $\sigma$ 
one finds  that the corresponding  flat-space partition function  
 that can be written  as 
 \be\la{155}
Z_{\conf,2} =   \Big[\frac{(\det\Delta_{0})^{3}}{\det\Delta_{ 2}}\Big]^{1/2}  = \Big[
\frac{\det \Delta_{1\perp}}{\det\Delta_{2\perp}} \Big]^{1/2} \Big[ \Big(\frac{\det\Delta_{0}}{\det\Delta_{1\perp}}\Big)^{2}
\Big]^{1/2}\ .
\ee
This is  recognised to be   a    product of  the standard  massless  spin 2  and  the  square   of  the massless spin 1 partition functions
(cf. \rf{1}). 

A curved space generalization of the equations  (\ref{1.2}) was first 
attempted in \cite{Grishchuk:1980hb}, but there the transversality constraint  
 $\na^{\mu}\,\varphi_{\mu\nu}=0$   was added by hand (i.e.  $V^\perp_\mu$ part  was  ignored) making  the theory effectively 
 non-local or non-Lagrangian.  
It was    pointed  out in  \cite{Grishchuk:1980hb}    that  the coefficient $\omega$ of the Weyl tensor  coupling in the 
generalization of \rf{1.2} 
 is, in general, arbitrary, 
being allowed  by the Weyl covariance condition. 
The consistent  $d=4$ Lagrangian \rf{1.10} 
first  appeared for  $\omega=0$ in  \cite{Deser:1983mm}.\foot{In general dimension $d$, its Weyl invariant 2-derivative
analog 
 found 
  in  \cite{Nepomechie:1983yq}  lacks scalar gauge invariance for $d >4$.} Its  $\omega=-2$  version was  
  found   in \cite{Leonovich:1984cf}.\footnote{The 
  sign of the curvature in \cite{Deser:1983mm,Nepomechie:1983yq,Leonovich:1984cf} 
was  opposite to the conventional one we use here:  $R^{\mu}{}_{\nu\rho\sigma} = \partial_{\rho}\Gamma^{\mu}_{\nu\sigma}+...$.
Note also that  ref.\cite{Gusynin:1986kz}  contained (up to a  misprint  in Eq.~(27))  the  conformal spin 2 operator  that is  the same as in  
\cite{Deser:1983mm}. 
Finally, there were  also attempts to define   the conformally invariant  operator on
a trace-full  rank 2  tensor $h_{\mu\nu}$ \cite{Achour:2013afa,Faci:2011dd}  which   we will not consider here 
as we are interested in irreducible representations of the Lorentz group.}
In its  general   form   (with an arbitrary $\omega$)  the Lagrangian 
 \rf{1.10} first appeared in  \cite{Erdmenger:1997wy}.

\subsubsection{Scalar gauge invariance  and partition function in  Ricci-flat space} 

An important   observation  made  in  \cite{Deser:1983mm} is that   restricted to a  conformally 
flat   homogeneous  background  ($AdS_4$  or  $S^{4}$)  the Lagrangian (\ref{1.10}) 
admits an analog  of the scalar gauge invariance  in \rf{1.3},
\be
\label{1.11}\te 
\delta \varphi_{\mu\nu} = \big(\na_{\mu}\,\na_{\nu}-\frac{1}{4}\,g_{\mu\nu}\,\na^{2}\big)\,\sigma \  .
\ee
This  curved-space generalization of the   scalar gauge invariance \rf{1.3}   is an important consistency requirement. 
As we shall explain in Appendix A,    this   invariance generalises    also to the Einstein-space backgrounds {\it provided} 
the    coefficient $\omega$  in \rf{1.10} is chosen to  be 
\be \la{111}  \om  =-2   \ ,  \ee
i.e. is the same   as in the  expansion of the Einstein action \rf{110} 
or in  the  Lichnerowicz  operator.\foot{Note that  ref.\cite{Erdmenger:1997wy} 
 considered a different  special value:  $\omega=-1$ in  $d=4$. For this choice, 
the Lagrangian (\ref{1.10}) 
may be written  in a factorized form 

$
\mathscr L = \frac{2}{3}\, (\mathscr D\varphi)^{\mu\nu\lambda}\,(\mathscr D\varphi)_{\mu\nu\lambda}, \qquad
(\mathscr D\varphi)_{\mu\nu\lambda} = \na_{\lambda}\varphi_{\mu\nu}-\na_{(\mu} \varphi_{\nu)\lambda}
-\frac{1}{3}\Big(
g_{\lambda (\mu} \na^{\rho}\varphi_{\nu)\rho}-2\,g_{\mu\nu}\,\na^{\rho}\varphi_{\lambda\rho}
\Big)\ .
$

The significance of this observation  is unclear, given that the  absence of  scalar   gauge  invariance
in this case. 
} 
The presence of the scalar gauge invariance in a gravitational background indicates that one   may be able to couple 
this  conformal  rank 2 tensor  field to Einstein gravity in a consistent way  (cf. \ci{Deser:2013bs}).

This  choice  is  special also   for another (related) reason:  in this case  the   differential  operator in \rf{1.10} 
factorises  on an Einstein  background:  the spin 2 and spin 1 transverse parts  in the  curved space   analog of the 
 decomposition \rf{1.5}  decouple  in the action. 
 This  makes the $\na^2$  kinetic operator effectively diagonal  and    allows  one to  write the curved space  generalization  of the partition function  \rf{155}   in a simple  form  in terms of determinants of the standard Lichnerowicz  spin $s=2,1,0$ operators. 
 
 Indeed, let  us consider  first   the Ricci-flat  background.  
 Using 
the covariant version of the decomposition \rf{1.5} 
\be
\label{2.14}\te 
\varphi_{\mu\nu} = \varphi_{\mu\nu}^{\perp}+\na_{(\mu}\,V_{\nu)}^{\perp}+ \big(\na_{\mu}\na_{\nu}-\frac{1}{4}\,g_{\mu\nu}\,\na^{2}\big)\,\sigma
\ee
 in \rf{1.10}   with $R_{\m\n}=0$ and $\om=-2$ we get 
 \ba 
  \mathscr  L_{\conf,2} 
&\te  = \na^{\lambda}\varphi^{\mu\nu}\,\na_{\lambda}\varphi_{\mu\nu}
-\frac{4}{3}\, (\na_{\m}\varphi^{\mu\n} )^2  -2 \,C_{\mu\nu\rho\lambda}\,\varphi^{\mu\rho}\,\varphi^{\nu\lambda} \no \\
& \te = 
\varphi^{\perp}_{\mu\nu}\, \Delta_{\LL\, 2}\,\varphi^{\perp\, \mu\nu} 
+ \frac{2}{3} V^{\perp}_{\mu} \, (\Delta_{\LL\, 1})^{2}\,V^{\perp\, \mu}  \ , \la{412} 
 \end{align} 
 where $ \Delta_{\LL\, s}$ are the  Lichnerowicz operators \rf{55}    on Ricci  flat background. 
 As was  mentioned above,   the decoupling of $\sigma$  (which is a  manifestation of the 
  scalar gauge invariance) and   separation of $\vp^\perp_{\mu\nu}$ and $V^\perp_{\mu}$ 
 is the consequence of  the choice  of $\om$ in \rf{111}.\foot{The reason for this   can be understood 
 by comparing to the case of the Einstein theory  \rf{110}  on $R_{\m\n}=0$    background 
  where the presence of the  $\om=-2$ 
  Lichnerowicz operator  is directly related to the  vector-parameter     gauge  invariance,  i.e. to 
   the  decoupling of the "longitudinal" part of 
  $h_{\m\n}$   before gauge fixing. }

 The  Jacobian  of the transformation in \rf{2.14} 
 is   $J=[ \det\Delta_{\LL\, 1\, \perp}\  (\det\Delta_{0})^{2}]^{1/2}$,  
so  we get the following generalization of the flat-space expression \rf{155} 
 \be
\label{2.21}
Z_{\conf,2}  =\Big[ \frac{ (\det\Delta_{0})^{2} }
{\det\Delta_{\LL\,2\,\perp}\,\det\Delta_{\LL\,1\, \perp}}\Big]^{1/2} = 
\Big[\frac{(\det\Delta_{0})^{3}}{\det\Delta_{\LL\, 2}}\Big]^{1/2} \ , 
\ee
where $(\Delta_{\LL\, 2})_{\mu\nu,\alpha\beta} = -  g_{\m (\alpha} g_{\beta)\n}\na^{2} -2 C_{\mu\alpha\nu\beta}$,
\  $(\Delta_{\LL\, 1})_{\mu\nu} = - g_{\m\n} \na^{2} $ and $\Delta_0 = - \na^2$. 


\subsubsection{Partition function on $S^{4}$}

Let us now compute the  partition function  corresponding to the action \rf{1.10}  on conformally flat Einstein background, e.g., $S^4$. 
Using \rf{2.14}  
we find that    $\sigma$ decouples (manifesting  scalar     gauge invariance) 
and the  dependence $\vp^\perp$ and $V^\perp$  separates  (we consider  unit-radius $S^4$, i.e.  $R=12$)
\be\la{la2}\te 
\mathscr L _{\conf,2} (S^4) = \varphi^{\perp}_{\mu\nu}\,\Delta_{2\perp}(4)
\,\varphi^{\perp\, \mu\nu}+
\frac{2}{3}\,V_{\mu}^{\perp}\,\Delta_{1\perp}(3)\Delta_{1\perp}(-3)\,V^{\perp\, \mu}\ .
\ee
As in \rf{7}--\rf{9}   here $\Delta_{s\perp}(M^{2}) = -\na^{2}+M^{2}$, acting in transverse symmetric traceless tensors of rank 
$s$.
Taking into account the Jacobian  of the transformation \rf{2.14} (see, e.g.,  (A.8) in  \cite{Tseytlin:2013jya})
\be
\label{2.16}
J = \Big[\det\Delta_{1\perp}(-3)\,\det\Delta_{0}(-4)\,\det\Delta_{0}(0)\Big]^{1/2},
\ee
we end with the following partition function (generalising  \rf{155} in the flat-space case where  all mass terms 
are  zero) 
\ba
\label{2.17}
&Z_{\conf,2} = 
  Z_{2,0}\ Z_{1,0}  \ , \qquad \qquad 
Z_{1,0} =  \Big[\frac{\det\Delta_{0}(0) } {\det\Delta_{1\perp}(3)}\Big]^{1/2} =
   \Big[\frac{(\det\Delta_{0}(0))^2 } {\det\Delta_{1}(3)}\Big]^{1/2}  \ , \\ & \ \ \ \ \ \ 
Z_{2,0}  = \Big[\frac{\det\Delta_{0}(-4)}
{\det\Delta_{2\perp}(4)} \Big]^{1/2} = \Big[\frac{\det\Delta_{0}(-4)\,\det\Delta_{1}(-1)}
{\det\Delta_{2}(4)}\Big]^{1/2} \ . \la{217}
\end{align}
Here $Z_{n,k}$   are defined as in \rf{7},\rf{8}, i.e. 
$Z_{1,0}$ is the standard Maxwell partition function and $Z_{2,0}$ is the partition function  corresponding 
to the 
 $s=2$ partially massless  field \ci{Deser:1983mm}.  $Z_{2,0}$ appears, together   with the massless  spin 2 (Einstein
  graviton)  partition function $Z_{2,1}$,    in the Weyl graviton partition function $Z_{\chs,2}= Z_{2,1} Z_{2,0}$
  which is  a special case of \rf{22} (see \cite{Tseytlin:2013jya}).
  Thus we  have the following relation 
  \be \la{712} 
  Z_{\conf,2} = {  Z_{\chs,2} \ Z_{1,0} \ov Z_{2,1}  } \ . \ee 
  In Appendix  B we shall also   present the expression for the
  finite temperature    partition function on $S^1_\beta \times S^3$  and discuss its interpretation 
  in terms of counting of conformal operators  in a  spin  2 CFT  in $\mathbb R^4$
  corresponding to   "shortened" representation  with shadow counterpart   as 
  $ (3; 1,1) - (5; 0,0)$.

\subsubsection{Conformal anomaly coefficients and AdS/CFT interpretation}

In $d=4$  the conformal anomaly coefficients $\aa$ and $\cc$ 
 in 
 \be\la{333} 
\te   b_4 = \b_1  R^*R^*   +\b_2   \big(R^2_{\m\n} - { 1 \ov 3} R^2\big) = -\aa\,  R^* R^* + \cc\,  C^2 \ , \qquad 
  \beta_1 = \cc- \aa \ , \ \ \ \ \   \beta_2 = 2 \cc \ , 
   \ee 
   can be   found by doing two separate computations: (i) on conformally-flat space like $S^4$ (finding $\aa$ coefficient) 
    and (ii) on a  Ricci-flat space  (finding $\cc-\aa$ coefficient). 
    The $\aa$ coefficient   corresponding to \rf{1.10} thus does not depend on $\omega$   and  is readily obtained  from 
    the  expression for the partition function in \rf{2.17}.
    Using that \cite{Tseytlin:2013jya}
    \be
\label{5.3}
\begin{split}
{\rm a}[\Delta_s(M^2)] &= \te   \frac{1}{144}  (s+1)^{2}) \Big[(s+1)^{2}-3M^{4}+12M^{2}-\frac{63}{5}\Big], \\
{\rm a}[\Delta_{\perp\, s}(M^2)] &=\te    \frac{1}{720}(2s+1) \Big[30s^{3}+85s^{2}+10s-58-30(s^{2}-2)M^{2}-15M^{4}\Big] \ , 
\end{split}
\ee
we get 
    \be \la{334}
  \te   \aa_{\conf,2} =  \aa_{2,0}+  \aa_{1,0} =   { 53\ov 45} +         {31 \ov 180} = {27\ov 20} \ . \ee
   At the same time, computing  the conformal anomaly  $\cc$  coefficient   for the  action  in \rf{1.10}  
   on a Ricci-flat background  is, in general,  a  non-trivial 
  problem. The reason is that  the corresponding second order operator  is a non-minimal  one, i.e.  has  "non-diagonal"   highest derivative part
 (cf.   \ci{Moss:2013cba}).  This problem  is, however,   readily solved   in the  special  $\omega=-2$ case \rf{111} 
 where one finds the explicit  factorized expression for the partition function \rf{2.21}. 
    Using that \cite{Christensen:1978md,Tseytlin:2013jya} 
    \be 
  \te   \b_1[\Delta_{{\rm L}\, s}] = {1 \ov 720} (s+1)^2  \big[  21 - 20 (s+1)^2  +   3 (s+1)^4  \big] \ , \la{335} \ee
    we find
    \be 
   \te  (\b_1)_{\conf,2}= {31\ov 30} \ , \ \ \ \ \ \ \ \ \ \ {\rm i.e.} \ \ \ \ \ \ \ \ \  \  \   \cc_{\conf,2}= { 143 \ov 60}   \ . \la{336}  \ee 
    Let us now   explain   how these  results  fit  into the general  expressions   for the conformal   anomaly coefficients 
    of  massive  $SO(2,4)$    representations  discussed in  \ci{Beccaria:2014xda}.
    Given a 4d conformal   field  in  representation $(\Delta'; j_1, j_2)$  the corresponding shadow field  
    $(\Delta; j_1, j_2)$  with $\Delta= 4 - \Delta'$   is  associated  (as a boundary value) to a massive   higher spin in $AdS_5$
    with  mass $m^2 = (\Delta-2)^2 - s^2, \ \ s= j_1 + j_2 $  \  \ci{Metsaev:2003cu} (the corresponding  kinetic operator  for $j_1 \geq j_2$  is 
    $-\nabla^2 + \Delta (\Delta-4) - 2 j_1$). 
     A non-unitary representation field  in 4d corresponds to a  ``partially massless" field in 5d. 
    The CST  field $(1; 1,1)$  is  thus  related  to   $(3; 1,1)$   spin 2 field in $AdS_5$. 
    In the  case  of  $(\Delta; {s\ov 2} , {s\ov 2})$  massive  field  one  finds    via $AdS_5$     computation \ci{Giombi:2013yva} 
    (as in \ci{Beccaria:2014xda} we use hat to distinguish a  long  representation from shortened one) 
    \be \la{63}
  \te  \hat  \aa(\Delta; {s\ov 2} , {s\ov 2})= { 1 \ov 720}  (s+1)^2  (\Delta-2)^3  \big( - 3 \Delta^2 + 12 \Delta + 5 s^2   +10 s -7\big) \ . 
    \ee 
    In the present case, accounting for the scalar gauge invariance, we  should  get  (see   \rf{777}; 
     here $\Delta'=1$ for the rank 2 field and $\Delta'=-1$ for the scalar gauge  parameter field) 
    \be  \la{644} 
  \te    \aa_{\conf,2} = \hat \aa(3; 1,1) - \hat   \aa(5; 0,0) =  { 27 \ov 20} \ .  \ee
    This  is indeed  in agreement with  direct computation in \rf{334}.

    The expression for $\b_1=\cc-\aa$ coefficient  for a conformal field  associated to a representation
   $(\Delta; {s\ov 2} , {s\ov 2})$   field in $AdS_5$ suggested in 
  \ci{Beccaria:2014xda}  was 
   \be \la{633}
 \te   \hat  \b_1(\Delta; {s\ov 2} , {s\ov 2})
   = { 1 \ov 720}  (s+1)^2  (\Delta-2) \Big[
    - 3 (\Delta-2)^4  - 5 ( s^2 + 2 s - 3)  (\Delta-2)^2     + 8 s^3   + 2 s^2   -12 s  -8 \Big] 
    \ee 
  Then  using also \rf{644}   we get 
  \be  \la{6444} 
\te      \cc_{\conf,2} = \hat \cc(3; 1,1) - \hat   \cc(5; 0,0) =  { 143 \ov 60} \ ,    \ee
  in agreement    with \rf{336}.  The  equivalent  result  is found 
   from the   expression for $\b_1$ proposed in \ci{Mansfield:2003gs,Ardehali:2013gra} 
   \be \la{646} 
\te   \hat  \b_1'(\Delta; {s\ov 2} , {s\ov 2})
    =   { 1 \ov 180}  (s+1)^2  (\Delta-2)      \Big( 1 +   { 1 \ov 4}  s (s +2)  \big[   3 s  ( s + 2 )  -14\big] \Big)    \ ,  \ee 
    i.e.   the   present $s=2$  case does not distinguish between the two  expressions for  $\b_1$ or $\cc$    discussed  in 
   \ci{Beccaria:2014xda}.

\subsection{General rank  $s$  case}
Let us now  consider the  general $s$   case with $d=4$ Lagrangian ${\cal L}_{\conf,s} $ in  \rf{1.7}. 
 It turns out   that  for  $s>2$ the  flat   space    scalar  gauge invariance in  \rf{117} 
 survives  in  curved space  action  \rf{1.7}  only in  the conformally-flat case $C_{\m\n\l\r}=0$,\footnote{Scalar gauge
invariance is expected on a 
four dimensional 
conformally flat background   as   it is present in
flat space  and the action in curved  background  is Weyl invariant (for a related observation for partially massless   fields in dS  space see \ci{Deser:2004ji}). } 
 i.e. in contrast to the $s \leq 2$ cases   it  is absent in the Ricci flat   background   for any 
 value of $\om$.\footnote{One  may  draw   an   analogy with 
 the  standard massless  spin $s$ field  with $\del^2$ Lagrangian. In flat space  one has maximal gauge invariance  with spin $s-1$   parameter. This  gauge invariance generalises to curved   space in the case   of conformally-flat  homogeneous 
 background.  However,   massless spin $s>2$  quadratic  actions    do not admit  consistent  gauge-invariant  generalizations 
 to   Ricci flat   backgrounds.  This is repaired in Fradkin-Vasiliev  type theory
where one adds all  spins together and cancels    variation under
gauge invariance by variation of graviton and   other  spins.  No such   consistent interacting 
theory   is known in the present  second  derivative  conformal  higher spin case   \rf{1.7},
but one may conjecture that it may exist.} 

Also, the     Weyl-covariant differential   operator in \rf{1.7}  does not factorize, i.e. is non-minimal
(has non-diagonal highest-derivative part in the transverse decomposition like \rf{1.5})  
unless the space is conformally flat.  
That  makes it hard to compute the corresponding conformal anomaly $\cc$-coefficient.


Starting with the flat space case, 
 the  partition function  corresponding to    \rf{117}   is  (cf. \rf{1},\rf{12} 
 \be \la{23} 
 Z_{{\conf}, s} =   \Big[ \frac{(\det\Delta_{0})^{s+1} }{\det\Delta_{s}}   \Big]^{1/2} 
 =  \prod_{k=1}^{s} \Big[
\frac{\det\Delta_{0}}{\det\Delta_{k\, \perp}}
\Big]^{1/2} =   (Z_0)^{\nu_s} \ , 
\ee
 where   the    corresponding number of   dynamical degrees of freedom is 
 \be \la{24} 
\te  \n_s = N_s - (s+1)  = \Big[\frac{(2 s + d -2) (s+ d-3)!}{(d-2)! s!} - s-1\Big] 
 _{d=4} = s(s+1)   \ . 
\ee 
Thus   the  number of degrees of freedom of CST   field  happens to be the same as for CHS  field   of the same spin  \rf{15} 
and thus   the corresponding  flat-space  partition functions  match.  
As a result,  
\be \la{249}
\big( Z_{{\conf}} \big)_\tot = \prod_{s=1} Z_{{\conf}, s}  =    1\ ,   \ee 
assuming one uses the same  regularization as in \rf{16}. 

Counting conformal   gauge-invariant  operators corresponding to the theory  \rf{117}  one  should be able to find the corresponding 
one-particle   partition  function generalising the $s=2$   expression in \rf{345}. The same   result should follow from the computation 
of the partition function on conformally-flat $S^1_\b \times S^3$   background. The $s > 2$ generalization of \rf{2.23} has the structure 
\be\la{888}
Z_{{\conf}, s}= \Big[\prod_{k=1}^{s}\frac{1}{(\det\bDelta_{k\, \perp})^{s-k+1}}\Big]^{1/2}  = 
\exp \Big[    \sum_{m=1}^\infty {1 \ov m} \Z_{{\conf}, s}(q^m) \Big]   \ ,  
\ee
where $q= e^{-\b}$ and  $\bDelta_{k\, \perp}$ are defined on 3d tensors  appearing in decomposition of $\vp_s$. 
Using the explicit spectra (and  omitting    zero modes  for fields   appearing in transverse decompositions  as in  \ci{Beccaria:2014jxa})
leads to the  following one-particle partition function
\ba
\no 
\Z_{{\conf}, s}(q) &= \sum_{k=1}^{s}\sum_{r=1}^{s-k+1}\sum_{n=r-1}^{\infty}
\,2\,(n+1)\,(n+2k+1)\,q^{n+s-2r+3} \   \\  \la{4.7} 
&= \frac{2 q^2 [ q^{s+1}-(s+1)^{2}\,q+s(s+2)]}{(1-q)^{4}} =   \frac{2 q^2 [ s(s+2) - q - q^2 - ... - q^s ] }{(1-q)^{3}}  \ .
\end{align}
This is the sum of the   contributions  of  spin $k=1, ..., s$   factors in \rf{888} 
\be
\begin{split}
k=s: & \qquad q^{n+s+1}, \ \ n\ge 0; \\
k=s-1: & \qquad q^{n+s+1},\ \  n\ge 0,  \qquad q^{n+s-1}, \ \ n\ge 1\\
k=s-2: & \qquad q^{n+s+1}, \ \ n\ge 0,  \qquad q^{n+s-1}, \ \ n\ge 1,  \qquad q^{n+s-3}, \ \ n\ge 2;  \ \ \ \   {\rm etc.}
\end{split}
\ee
The  Casimir energy on $S^3$   can be found the  $\beta\to 0 $ expansion of the one-particle partition function as 
$
\Z(e^{-\beta}) = \text{poles}+\text{constant}-2\, E\beta+\mc O(\beta^{2})$. From \rf{4.7} we get     
\be\la{eec}
E_{{\conf}, s} =\te   \frac{1}{720} s (s+1) (6 s^3+24 s^2+16 s-13) \ .
\ee
Summing this  over $s$ with the cutoff \rf{16}  gives, in contrast to the CHS case,   a non-zero result ($\sum_{s=1}^\infty  E_{{\conf}, s} ={43 \ov 448}$).

The partition  function on $S^4$    background that generalises the low-spin expressions   in \rf{2.17}  and \rf{z3} 
may be written as  (cf. \rf{9},\rf{22})\foot{The "ghost"  (numerator)  factor   here 
$ \prod_{k=1}^{s}  \det\Delta_{0}(  M^2_{0,k}   ), \ \ M^2_{0,k} =   2 - k - k^2 $  is very similar to 
the  one appearing in the determinant  of the  Weyl-covariant $\na^{2s} + ... $  scalar  GJMS operator \ci{Graham:2007}
 $ \prod_{k=1}^{s}  \det\Delta_{0}(m^2_k )$, where $m^2_k = (\ha d -k ) ( \ha d + k-1)\Big|_{d=4} = 2 + k - k^2 = M^2_{0,k-1}  $.}
\ba\la{25} 
&Z_{{\conf}, s}(S^4) = \prod_{k=1}^{s} Z_{k,0}\ , \qquad \qquad 
Z_{k,0} = \Big[
\frac{\det\Delta_{0}( M^2_{0,k})}{\det\Delta_{k\, \perp}(M^2_{k,0})}
\Big]^{1/2}\ , \\   & \ \ \ \ \qquad \qquad 
M^2_{k,0} = 2 + k \ , \ \ \qquad    M^2_{0,k}= 2 -k-k^2 \ . \no 
\end{align}
The corresponding  logarithmic divergence coefficient or   conformal anomaly a-coefficient 
  can be found using \rf{5.3}
\be \la{26} 
{\rm a}_{{\conf}, s} = \sum_{k=1}^{s}\Big(
{\rm a}[\Delta_{k\, \perp}(2+k)] -{\rm a}[\Delta_{0}(2-k-k^{2})]
\Big) = \te  \frac{1}{720} s (s+1)^2 (3 s^2+14 s+14)\  .
\ee
Like the sum of Casimir energies \rf{eec} 
the sum  of  conformal anomalies   $\aa_\tot=\sum_{s=1}^\infty {\rm a}_{{\conf}, s}$  does not vanish   in any natural regularization
(in the regularization used in \rf{16} one gets $\aa_\tot=-{195\ov 896}$). 
There is, of course, no   a priori  reason why one needs to sum over all  ranks $s$  here: 
in the MHS   and CHS cases this  summation was   implied, in particular, 
 by the AdS/CFT duality and  a relation to conserved  currents  of the boundary  theory, 
but this   connection  is absent here. 


Since  the  CST  field  corresponds to the  $SO(2,4)$ representation \rf{777},  it  can be  associated to  a particular  
 field in  $AdS_5$  for the corresponding combination of shadow representations, 
 $(3; {s\ov 2}, {s\ov 2}) - ( 3+s;0,0)$.
 This   "maximal depth" conformal   field  has  scalar gauge invariance, i.e. can be interpreted as  corresponding to 
  "maximal-depth"  partially massless
 field  in    $AdS_5$  \ci{Bekaert:2013zya}.
  In general,  the   one-particle partition function  of 4d CFT 
   can then be written  in terms of $AdS_5$ partition functions or conformal characters as 
    (see \ci{Beccaria:2014jxa,Beccaria:2014xda})
\ba\la{az}
&\Z_{\conf,s}(q) = \Z^{-}_s(q) -\Z^{+}_s(q) = \Z^{+}(q^{-1})-\Z^{+}(q)+\sigma(q) \ ,  \\
&\Z^{+}_s(q) = \widehat\Z^{+}(3;\, \tfrac{s}{2}, \tfrac{s}{2})- \widehat\Z^{+}(s+3;\, 0,0)
= \frac{(s+1)^{2}\,q^{3}-q^{s+3}}{(1-q)^{4}} \ . \la{aaz} 
\end{align}
 $ \widehat\Z^{+}(\Delta; j_1,j_2)= ( 2 j_1 +1) (2 j_2+1)  { q^\Delta \ov (1-q)^4}$  is the character 
of a  massive (long)  conformal group  representation $(\Delta; j_1,j_2)$.  
 $\sigma(q)$ is a "correction term"   which is expected in the presence of gauge symmetry (scalar gauge invariance here); 
 it  is even in $q\to 1/q$ and absorbs poles in
 the  naive  combination $\Z^{+}_s(q^{-1})-\Z^{+}_s(q)$. 
 The  minimal choice 
\be \la{sig}
\sigma(q) ={\te  \frac{1}{6}}  s(s+1)(s+2)+  {\te \frac{1}{6}}  \sum_{n=1}^{s-1} n (n+1)(n+2)(q^{n-s}+q^{s-n}) 
\ee
leads to $\Z_{\conf,s}(q)$   in \rf{4.7}. 

In the case of $S^4$   background   the expression for the a-anomaly coefficient 
can be computed using the $AdS_5$ relation as in \ci{Giombi:2013yva,Beccaria:2014xda}, i.e.
 using \rf{63}   and   generalising  \rf{644} 
\be  \la{anm} 
\te {\rm a}_{{\conf}, s} =  \hat \aa(3; {s \ov 2}, {s\ov 2} ) - \hat   \aa(3+s; 0,0) =
   \frac{1}{720} s (s+1)^2 (3 s^2+14 s+14)\  .
\ee 
This agrees   with the  result of the direct computation in $d=4$   in \rf{26}.

As for the  c-anomaly coefficient,  its computation directly from the action \rf{1.7} with $s >2$ 
 on Ricci-flat background is problematic for two  reasons: 
(i)  the  lack of scalar gauge invariance making the resulting partition  function scalar gauge dependent;
(ii)  the non-minimal nature  (lack of  factorization) of the corresponding  second order differential operator 
requiring to use more complicated methods (cf. \ci{Moss:2013cba})  than the standard algorithm  for the $b_4$ Seeley   coefficient. 
If one   ignores these problems  one may  formally  generalize  the known $s=1$  and $s=2$  expressions \rf{2.21}    
 to the $s >2$ Ricci   flat case  in the way  directly analogous   to the flat space  case \rf{23} 
 (cf. also \rf{44} and \rf{18}) 
  \be\la{8888}
\td Z_{{\conf}, s} =
\Big[
\frac{(\det\Delta_{ 0})^{s+1}}{\det\Delta_{{\rm L}\,s} }\Big]^{1/2}\ .
\ee
Then \rf{335}  implies that  the corresponding $\beta_1 = \cc-\aa$   is given by 
\be 
(\td \b_1)_{{\conf}, s} = 
\b_1[\Delta_{{\rm L}\,s} ]- (s+1) \b_1[\Delta_{0} ]
\te =  \frac{1}{720} s (s+1) (3s^4 + 15 s^3  + 10 s^2 -30 s  -24 )  \ , \la{788}
\ee 
generalising \rf{336}.\foot{As in the  case of the  Casimir energy and $\aa$-anomaly,  summing \rf{778}  over $s$  does not give  vanishing result in any reasonable regularization.}

We may  compare  \rf{788}  with  prediction based on  dual  $AdS_5$  description.
If one uses  the expression  \rf{646} of  \ci{Mansfield:2003gs,Ardehali:2013gra}   one gets 
\be 
\te (\td \b_1)'_{{\conf}, s} 
= \b'_1(3; {s \ov 2}, {s\ov 2} ) - \b'_1(3+s; 0,0) 
=  \frac{1}{720} s (s+1) (3s^4 + 15 s^3  + 10 s^2 -30 s  -24 )   \ , \la{778} \ee 
i.e. the same result as in \rf{788}. 
At the same  time, the expression for $\b_1=\cc-\aa $ for representation $(\Delta; {s\ov 2},{s\ov 2}) $   suggested in  \ci{Beccaria:2014xda}
leads  to a different  value:
\be \la{hhh}
\te (\td \b_1)_{{\conf}, s} = (\td \b_1)'_{{\conf}, s}  -    \binom{s+3}{6} \ , \ \ \ \ \    \qquad   \binom{s+3}{6}=
 \frac{1}{720} (s-2)(s-1)s(s+1)(s+2)(s+3) \ . \ee 
The  two  results  agree   for $s=0,1,2$ but disagree  by an integer   for $s > 2$. 

It is interesting to note that a  similar   conclusion is reached   in the   conformal higher spin case.
Like  \rf{8888}   the factorized   expression \rf{18} for the CHS partition function on a generic Ricci-flat background 
 is  not  justified for $s >2$.\foot{One  may   conjecture   that  it   may  still apply  to the computation of the $\b_1$    anomaly
 coefficient. For example, terms   that obstruct factorization may depend only on derivatives of the curvature and  thus   do not contribute 
to  $\b_1$.}  Still, starting with \rf{18}   and  applying  \rf{646}  to the computation of $(\beta_1)_{\chs,s}$    one gets the same expression as in \rf{199};
at  the same time, $\beta_1(\Delta; {s\ov 2}, {s\ov 2})$   or \rf{63}   suggested in  \ci{Beccaria:2014xda}
gives the result   differing by the same integer as in \rf{hhh}\foot{Note that $ \sum_{s=1}^\infty   \binom{s+3}{6}\ e^{-\ep(s+ {1\ov 2} )} \Big|_{\rm fin} =0$,   so   both expressions   for $\beta_1$ are consistent with  the vanishing of the sum of $\cc_s$ over $s$ (see remark below  \rf{199}).}
\be \la{hhhh}
\te (\b_1)_{{\chs}, s} = (\td \b_1)'_{{\chs}, s}  -    \binom{s+3}{6} \  . 
\ee
The   significance of this observation is not clear at the moment.\foot{Let us add also   that  
analogous   conclusion applies to the    scalar GJMS operator discussed in appendix A.1.
For example, 
 in $d=4$   the corresponding value of the $\aa$-coefficient that follows from \rf{d1},\rf{d7}  is 
$\aa_r = - { 1 \ov 720}  r^3 ( 3 r^2 -5)$,  in agreement   with the 
$AdS_5$ expectation \rf{63}  for the representation $( 2 +r; 0, 0)$
(the scalar field has dimension $2-r$). 
The form of  this operator  on Ricci-flat background is
 not known explicitly    for $r > 4$ (cf. \ci{Gover:2005mn,Manvelyan:2006bk,Juhl:2009aa,Juhl:2011aa,Fefferman:2012aa}). 
The  $r=3$  operator  is singular in $d=4$ (the singular term is proportional to Bach tensor).
Assuming   that  GJMS operator in $d=4$  extended to any $r$  (i.e. beyond the "critical" order $r_c=\ha d$)
  factorises on Ricci-flat background   becoming  $(\na^2)^r$, 
 the corresponding $\beta_1$ coefficient   is readily found to be $r$ times   the standard scalar one, i.e. 
 $\b'_1 =  { 1 \ov 180} r$. This is the same  result that follows from \rf{646}  for the representation $( 2+r; 0,0)$. 
 At the same  time, from \rf{646} we get $\b_1= { 1 \ov 720} r ( - 3r^4 + 15 r^2 -8)$. Thus  here 
 $\b_1 = \b_1'  - \ha  \binom{r+2}{5}$, which is similar to \rf{hhh} and \rf{hhhh}. 
  }



\section*{Acknowledgments}
We  thank  M. Grigoriev,   E. Joung, R. Metsaev and D. Toms  for useful discussions. 
This work    is  a part of  collaboration supported 
 by  the  Russian Science Foundation grant 14-42-00047 and associated with Lebedev Institute.
The work of AAT was   supported  also by   the  
ERC Advanced grant No.290456.  
AAT  would like also to thank the Galileo Galilei Institute for Theoretical Physics   for the hospitality   and the INFN for partial  support   
  while part of  this work was in progress. 



\appendix

\section{Relation  between   
partion functions on $S^4$ and 
on $AdS_5$       }
\label{appa}

The aim of  this Appendix is to  consider  in detail the   relation \rf{20}   by computing the determinants involved. 
As discussed in sections 2.1.3, 2.2.3  one   should carefully  correlate the UV regularization in 4d and 
 IR regularization  in 5d.  We shall follow  the 
 approach  of \cite{Diaz:2007an} where dimensional regularisation $4\to d=4-\varepsilon$
 was used  in demonstrating a   similar relation  in  spin 0 case. 
  $\varepsilon\to 0 $ plays the role of a common  
IR/UV regulator on the  AdS/CFT sides. 
  This allows  a careful separation between the divergent   pole terms $\sim \frac{1}{\varepsilon}$
and  finite remainder,   accounting, in particular,  for the IR finite terms  in  the $AdS_5$ partition function 
ignored in  direct  zeta-function regularization  in  \ci{Giombi:2014iua}. 
Matching the finite terms  turns out to  be quite   subtle. We  
 will  first  illustrate  this   on the  example of the partition function of  GJMS operators  on $S^{4}$
extending the analysis of \cite{Diaz:2007an,Diaz:2008hy}. 
We shall  then   turn  to the  more complicated   but structurally similar  
case of  the CHS  fields on $S^{4}$ related to MHS  fields on $AdS_5$.

\subsection{Matching partition function of  GJMS operators on $S^4$ and ratio of scalar  partition functions on  $AdS_5$  }

GJMS operators \cite{Graham:2007} are the unique  Weyl-covariant $\nabla^{2r} + ...$  operators in 
$d$   dimensions defined on scalars. 
In $d=4$ the $r=1$ case is familiar $-\nabla^2 + {1\ov 6} R$ and the $r=2$ operator 
 $\nabla^4+...$ was constructed in \ci{Fradkin:1981jc,Paneitz:1983}. 
Their general properties are discussed 
in \cite{Gover:2005mn,Manvelyan:2006bk,Juhl:2009aa,Juhl:2011aa,Fefferman:2012aa}. 
On an Einstein space 
   background, they factorize into a product  of  $r$ scalar   2nd-order operators 
\be
\mathscr{D}_{(2r)} = \prod_{k=1}^{r}\big(-\nabla^{2}+q_{k} R\big) \ , \qquad  \ \ \ \ \ \ \ \      q_k= \frac{(\frac{d}{2}-k)(\frac{d}{2}+k-1)}{d(d-1)} \ . \la{d1}
\ee
Using that on  a  unit sphere $S^{d}$  (with $R=d(d-1)$)  the  eigenvalues and multiplicities of $-\nabla^2 + M^2$  
are 
\be \lambda_n= n(n+d-1) + M^2 \ , \ \ \ \ \qquad    \dd_n=  \frac{(2n+d-1)\,\Gamma(n+d-1)}{\Gamma(d)\Gamma(n+1)}\ , 
\la{d2} \ee 
we get explicitly\foot{As discussed in \cite{Diaz:2007an}, in dimensional regularisation one has  a useful  relation
$\sum_{n=0}^{\infty}{\rm d}_{n}=0$  allowing to drop  constants under  the logarithm of  egenvalues.}
\be
\label{D.2}
\ln\det\mathscr D_{(2r)} = \sum_{n=0}^{\infty} \frac{(d+2n-1)\,\Gamma(d+n-1)}{\Gamma(d)\Gamma(n+1)}\ 
\ln \Big[\prod_{k=1}^{r}(n+k+\frac{d}{2}-1)(n-k+\frac{d}{2}) \Big] \ .
\ee 
From (\ref{D.2}), it is clear  that  the critical order $r=r_{c}\equiv \tfrac{d}{2}$ is special:  (i) 
for  $r<r_{c}$  all eigenvalues
are positive; (ii)  for  $r=r_{c}$  there is one zero mode; (iii)  for $r>r_{c}$ there are zero and negative eigenvalues
(see also \cite{Dowker:2010qy}). 

GJMS   theory   may be  viewed as an induced conformal theory on $S^{d} $ boundary of ${AdS}_{d+1}$  corresponding to a standard 2nd-derivative scalar  in $AdS_{d+1}$  and   the analogue of \rf{20}  reads\foot{In this Appendix we  add  hats to Laplacian operators 
not to confuse them with dimension parameter  $\Delta$.}  
\ba\la{d3}
&  \qquad \qquad Z_{(2r)}(S^{d})= \frac{Z^{-}_{0}(\text{AdS}_{d+1})}{Z^{+}_{0}(\text{AdS}_{d+1})} \ ,    \\
& Z_{(2r)} \equiv  \big( \det \mathscr{D}_{(2r)}\big)^{-1/2}  \ ,   \ \ \  \ \qquad \ \    
      Z_0 = \big( \det  \hDelta_0  \big)^{-1/2}   \ ,  \  \ \ \ \ \ \  \hDelta_0 = - \nabla^2 + m^2 \ ,  \la{d33} 
\end{align}
where 
$Z^{\pm}_{0}$ is the partition function of a  massive scalar operator   in $AdS_{d+1}$ with 
$m^2= \Delta(\Delta- d)=r^2 - {d^2 \ov 4}$, i.e. with  the associated operators having dimensions $\Delta_+ = { d \ov 2} + r$ and 
$\Delta_- = { d \ov 2} - r$. \footnote{Let us note that   \cite{Diaz:2007an}  considered  the case  of a generic massive $AdS_{d+1}$ scalar  with non-integer $r= \sqrt{ m^2 + {d^2 \ov 4}}$
 when the associated induced  boundary theory is non-local: the kinetic 
operator is  inverse of $K(x,x')=  < J(x) J(x') > = { [s(x,x')]^{-2 \Delta_-}}$ where $s(x,x')$    is the  geodesic distance on  (conformally) flat space.
Then the l.h.s. of \rf{d3} is  replaced by the partition function of  the corresponding non-local operator 
which may be interpreted  in terms of 
a double trace deformation of CFT   corresponding  to  the change of boundary conditions for the dual $AdS_{d+1}$  theory 
  \cite{Gubser:2002vv,Hartman:2006dy}. 
 For  an  integer $r$  the  non-local  kinetic operator ${ [s(x,x')]^{-2 \Delta_+}}$   has the leading   singular  being the 
 local  GJMS  operator acting on a delta-function (see also \ci{Diaz:2008hy,Tseytlin:2013jya}); thus   
  the  inverse of the determinant  of $K(x,x')$   is   effectively replaced by  the determinant of the  local conformally-covariant 
  GJMS operator as  in \rf{d3}. 
}

We  may now follow the procedure in   \cite{Diaz:2007an} to evaluate (\ref{D.2})
by first   using that 
\be
\prod_{k=1}^{r}(n+k+\frac{d}{2}-1)(n-k+\frac{d}{2}) = 
\frac{\Gamma(n+\frac{d}{2}+r)}{\Gamma(n+\frac{d}{2}-r)} \ ,
\ee
and  also    replacing   $r$   by    $\Delta \equiv \Delta_+= { d \ov 2} +r$
(we shall   use the notation    $\mathscr D_{(2r)}\to \mathscr D(\Delta)$).     
 Formally  treating   $\Delta$ as  a continuous variable,  \rf{D.2}  becomes 
\be\la{d4} 
\ln \det\mathscr D(\Delta) = \sum_{n=0}^{\infty}
 \frac{(d+2n-1)\,\Gamma(d+n-1)}{\Gamma(d)\Gamma(n+1)}\ln 
\frac{\Gamma (n+\Delta )}{\Gamma(n+d-\Delta)}\ .
\ee
It is convenient to  first take  derivative of \rf{d4} with respect to $\Delta$, do the sum    and then   integrate 
over $\Delta$, fixing the integration constant by demanding that  the result   should vanish at $r=0$ or $\Delta = {d\ov2}$ 
when  the GJMS operator becomes  trivial. 
The sum 
\be
\frac{\partial}{\partial\Delta}\ln \det\mathscr D(\Delta) = \sum_{n=0}^{\infty}
 \frac{(d+2n-1)\,\Gamma(d+n-1)}{\Gamma(d)\Gamma(n+1)}
 \Big[
 \psi(n+\Delta)+\psi(n+d-\Delta)
 \Big].
 \ee
was already    computed  in   \cite{Diaz:2007an}: 
\be
\label{D.8}
\begin{split}
 \frac{\partial}{\partial\Delta}\ln \det\mathscr D(\Delta) = 
 \frac{  \Gamma(-\tfrac{d}{2})\,  \Gamma (d-1) (d-2 \Delta )  \Gamma(\Delta )  \Gamma (d-\Delta ) \,\sin \big[\tfrac{\pi }{2}  (d-2 \Delta )\big]}{ 2^{d}\sqrt{\pi } \, \Gamma (d-1) 
   \Gamma \big(\frac{d+1}{2}\big)}
  \ .
\end{split}
\ee
Let us   now  turn to the $AdS_{d+1}$ side. Here one can  express the  derivative of the r.h.s. of \rf{d3}  over the  scalar mass or $\Delta$    
in terms of an integral of the trace  of the corresponding   difference of the $AdS_{d+1}$  scalar 
   bulk-to-bulk  propagators  \cite{Diaz:2007an}. 
   In general, for any spin $s \geq 0$ 
one can  utilise the  expression in eq.(71) of \cite{Costa:2014kfa} 
 for the mass$^2$  derivative   of the difference of the   logarithms 
of  the  partition functions (or, equivalently, the  difference of the bulk-to-bulk propagators) 
 for  the  spin $s$ symmetric transverse traceless field
in $AdS_{d+1}$   with  the kinetic operator (cf. \rf{9}) 
$\hDelta_{s\, \perp} = - \nabla^2   +  m^2, \   m^2= \Delta ( \Delta - {d } ) - s $\ 
corresponding to the  standard  (+) and   alternative  (-)  boundary conditions
(or dimensions $\Delta_+=\Delta$ and $\Delta_-=d-\Delta$)\foot{More precisely, eq.(71) of \cite{Costa:2014kfa}   gives 
$\ha {\del \ov \del  m^2}  \ln  {\det_- \hDelta_{s\, \perp} \ov \det_+ \hDelta_{s\, \perp}} 
 =
[  4 (\Delta - { d \ov 2} ) ]^{-1} {\del\ov \del  \Delta }  \ln {\det_- \hDelta_{s\, \perp} \ov \det_+ \hDelta_{s\, \perp}} $
Note that here $m^2= M^2 \vab=-M^2$  in  the notation in \rf{7},\rf{9}.    }
 \be
 \label{D.9}
\begin{split}
&\frac{\partial}{\partial\Delta}\ln 
{\det_- \hDelta_{s\, \perp} \ov \det_+ \hDelta_{s\, \perp}}
=\pi^{\frac{d}{2}}\,\Gamma(-\tfrac{d}{2})\,  \frac{(2s+d-2) (\Delta-\tfrac{d}{2})\,\Gamma(s+d-2)}{\Gamma(d-1)\Gamma(s+1)}
 \\
& \qquad \times  \frac{(\Delta+s-1)(\Delta-s-d+1)\Gamma(\Delta-1)\Gamma(d-1-\Delta)\sin\big[\tfrac{\pi}{2}
(d-2\Delta)\big]}{2^{d-1}\,\pi^{\frac{d+1}{2}}\Gamma(\frac{d+1}{2})}\ .
\end{split}
\ee
In (\ref{D.9})   we separated  the factor of the 
 dimensionally regularised volume of  $AdS_{d+1}$  (cf. \rf{vev})  
\be
V_{d+1} = \pi^{\frac{d}{2}}\,\Gamma(-\tfrac{d}{2})\ .
\ee
Comparing (\ref{D.8}) with the  $s=0$ case of  (\ref{D.9}), we find
 \be
 \label{D.11}
\frac{\partial}{\partial\Delta}\ln \det\mathscr D(\Delta) = 
\frac{\partial}{\partial\Delta}\ln \frac{\det_-\hDelta_{0}}{\det_+ \hDelta_{0}}  \ , 
 \ee
 which is indeed   in  agreement with \rf{d3}. 
 
Expanding around $d=4$, i.e. setting  $d=4-\varepsilon$ with $  \va\to 0$, 
one can check that the relation (\ref{D.11}) holds for both the pole term and the 
finite remainder.   The expansion of (\ref{D.8})  is  
\be\la{faf}
\begin{split}
&\qquad \qquad \qquad \qquad\frac{\partial}{\partial\Delta}\ln \det\mathscr D(\Delta)= \frac{1}{\varepsilon} P(\Delta)  +   F(\Delta) 
 +\mc O(\varepsilon), \\
&\qquad \qquad
P(\Delta) = 
{\te -\frac{1}{6}} (\Delta -3) (\Delta -2)^2 (\Delta -1)     \   , \\
& \qquad \qquad
 F(\Delta)= 
\te \frac{1}{72} (\Delta -2) (\Delta -1) \Big[12 (\Delta -3) (\Delta -2)  \gamma_{\rm E} +(107-25 \Delta) \Delta \\
&\ \ \qquad   \qquad 
 -6 \pi  (\Delta -3) (\Delta -2) \cot (\pi  \Delta )
   +12 (\Delta -3) (\Delta
   -2) \psi ^{(0)}(3-\Delta )-108\Big]\ .
\end{split}
\ee
Let us  now integrate  \rf{D.11}  in the interval 
$
\frac{d}{2}\le \Delta \le \frac{d}{2}+r
$. 
The integral of the pole is 
\be
\label{D.15}
 \frac{1}{\varepsilon} \int_{d\ov 2}^{{d\ov 2}+r} d\Delta\   P(\Delta) 
= {\te 
- \frac{1}{90 } r^3 \,(3 r^2-5) } { 1 \ov \va} + {\te \frac{1}{12}} r^2
   (r^2-1)+\mc O(\varepsilon).
\ee
Here the  coefficient   of the 
 singular  part is in agreement (after taking into account normalizations) 
 with the  conformal anomaly  coefficient ${\rm a}_{r} = -\tfrac{1}{720}r^{3}(3r^{2}-5)$ for the GJMS field 
 that can be computed directly  from  the Seeley coefficients $B_4$  for the operators in \rf{d1}.

 The full finite part  of  $\ln \det\mathscr D(\Delta)$      is the sum of the second term in (\ref{D.15}) and 
the integral of the second term $F(\Delta)$ in \rf{faf} (here we  may set $d=4$, i.e. $\Delta = 2 + r$)   
\be \la{d5}
X (\Delta) = \int_{2}^{\Delta }d\Delta'   \  F(\Delta')  \ , \ \ \ \ \ \ \ \ \ \ \ \  \Delta= 2 + r \ . 
\ee
The function $X(\Delta)$ has poles at $\Delta=4, 5, 6, \dots$. 
These poles
are associated with the previously  mentioned  zero eigenvalues appearing in 
(\ref{D.2})  for $r\ge r_{c}= { d \ov 2}$.\footnote{In addition,  $X(\Delta)$  contains 
 imaginary terms (multiples of $\pi$) that are related to negative eigenvalues
in (\ref{D.2}) appearing  for $r > r_{c}$  (note  also that for  $r\ge r_{c}= { d \ov 2}$  one has $\Delta_- = {d\ov 2} - r < 0$).
 In the following, we shall  formally omit these imaginary  contributions
that appear   on both sides of \rf{d3}. 
}
The poles of $X(\Delta)$ inside
the interval $(2,2+r)$ can be evaluated by taking the principal part of the integral. 
The rightmost pole will give a singular term $\sim \ln \big[\Delta-(2+r)\big]$. 
As a result, we find for $d\to 4$,\  $\Delta \to  2 + r$: 
\ba\la{d6}  &
 \ln \det\mathscr D(\Delta)=  \Big( { 1 \ov \va} -    \gamma _{\rm E} \Big)    \PP(\Delta) + \FF(\Delta) 
 +\mc O(\varepsilon)\ , \\ &   \qquad \PP(2+r)  
 =8 \aa_r= \te  - \frac{1}{90} r^3 \,(3 r^2-5)\ ,  
\la{d7} \\
&  \te 
\FF(2+r-\delta)\Big|_{\delta \to 0} =
\frac{1}{3}\ln A-\frac{2}{3} \zeta '(-3)+\frac{1}{6}
   \big(r^2-1\big) r^2\,  \ln \Gamma(r)\no \\
&\quad \te 
+\frac{1}{3}
   \big(r-2 r^3\big) \psi ^{(-2)}(1-r)-\frac{1}{4 \pi
   ^2}\big(r^2-1\big) \zeta (3)+\big(\frac{1}{3}-2 r^2\big) \psi ^{(-3)}(1-r)\no \\
   &\quad\te 
   +\frac{1}{12}
   \big(r^2-r^4\big) \ln (2\pi )-4 r \psi ^{(-4)}(1-r)-4 \psi
   ^{(-5)}(1-r) \label{D.19}\\
   &\quad\te 
   +\frac{1}{2160}\big(-150
   r^5+45 r^4+130 r^3-90 r^2-22\big)
   +\frac{1}{12} 
   \big(r^2-1\big) r^2 \, \ln \delta +\mc O(\delta)\ , \no 
\end{align}
where  
$A$ is the Glaisher constant, i.e. $\ln A= - {1\ov 2 \pi^2} \zeta'(2)  + {1\ov 12} \ln (2\pi ) + {1\ov 12} \gamma_{\rm E}$.  
 As anticipated, the last term in (\ref{D.19}) is due to the zero modes appearing in the 
original expression (\ref{D.2}) when $r\geq r_c=2$ in  $d=4$. The prefactor of $\ln \delta$  is  indeed 
 the sum of the associated multiplicities.
When  the zero modes are projected out  and thus $ \ln \delta$ terms   are omitted,    (\ref{D.19})  is given 
by a finite   expression 
   containing various transcendental constants and logarithms of integers.
   The   expressions for the polylogarithms  in \rf{D.19}   can be put in a  more explicit form 
   so that we find  after dropping the $\ln \delta $ terms
   \be
\label{D.21}
\begin{split}
\FF(2+r) &= 
\te  \frac{1}{3} r \big(2 r^2-1\big)\ln A
-\frac{2 }{3}\, r\, \zeta'(-3) -\frac{1}{720} r \big(60 r^2-31\big)+\mc L_{r}\ , 
\end{split}
\ee
where $\mc L_{r}$ is a sum of logarithms of  integers  which   general 
dependence on $r$  we did not  find.\foot{Explicitly,  for low  values of $r=1,...,5$  we get 
$\mc L_{1,2,3}=0, \ \mc L_{4}= 2 \ln 2$, \ $ \mc L_{5}= 2 \ln 3 + 12 \ln 2$,  i.e.
\be\nonumber
\begin{split}
 \FF(3) &=
 \frac{\ln A}{3}-\frac{2}{3} \zeta
   '(-3)-\frac{29}{720}\ , \ \ \ \ \ 
 \FF(4) = 
  \frac{14 \ln A}{3}-\frac{4}{3} \zeta '(-3) -\frac{209}{360} \ , \ \ \ \ 
   \FF(5) =
   17 \ln A-2 \zeta '(-3) -\frac{509}{240}
   \ , \   \\
     \FF(6) &=
      \frac{124 \ln A}{3}-\frac{8}{3} \zeta '(-3)
  -\frac{929}{180} +2 \ln 2 
   \ ,  \ \ \ \qquad
  \FF(7)    = 
  \frac{245 \ln A}{3}-\frac{10}{3} \zeta '(-3)
      -\frac{1469}{144}+2 \ln 3+12 \ln 2 
   \ .
\end{split}
\ee
}

\subsection{Matching  CHS partition function  on $S^4$ and  ratio of MHS  partition functions on  $AdS_5$  }

Let us now  repeat the  above discussion  of  the   scalar  case   for the $s>0$   CHS theory on $S^4$  to   demonstrate  the relation \rf{20} 
to MHS partition function   on $AdS_5$. Once again, 
 one  should  use the dimensional regularization  that provides   UV  regularization on the CHS boundary   side  
and IR regularization on the MHS  bulk side   at the same time.  
The matching of the  logarithmically divergent parts  was already demonstrated in \ci{Giombi:2013yva,Tseytlin:2013jya}.
The use of dimensional   regularization is  essential  on $AdS_5$ side as otherwise   one misses  the  crucial finite part  
of  $\ln Z_\MHS$   that  is always present in 
$\ln Z_\CHS$. 
Below  we shall limit  our analysis  to  matching  the most transcendental 
parts of the finite contributions to the partition functions on the two sides of \rf{20}, 
i.e.  the terms  proportional to $\ln A$ and $\zeta'(-3)$   as in  \rf{D.21}.

The starting point  on the CHS side should  be the $d=4\to d=4-\va$ 
generalization (cf.    \rf{166}) of  the  partition  function \rf{22} on $S^4$.
As  we will be interested only in the most transcendental  finite part of 
\be   G_s   \equiv  -  \ln Z_{\CHS,s}  (S^4)   \ , \la{aa1} \ee
we may ignore the  analog of the  $d-4$ power factor  in \rf{166}  
(it will contribute only to rational finite  terms)
and  start   directly   with  \rf{22}   in $d=4$   introducing some  fiducial UV cutoff
(the most transcendental terms   in \rf{aa1}   should not depend on its choice, cf. \cite{Diaz:2008hy}).

  Using the  known spectra of $\Delta_{s\, \perp}$   operators  on $S^4$ (see, e.g., 
\ci{Tseytlin:2013fca})  we find from   \rf{22} 
\ba
\label{D.22}
  G_s  &= 
  \sum_{n}^{\L-s-2}\, \sum_{k=0}^{s-1} Q_{s}(n,k) \ , \ \ \ \ \ \ \ \ \  \ \ \ \ \ \   \L \to \infty \\
 Q_{s}(n,k) &=\te \frac{1}{12} (n+1) (2 s+1) (n+2 s+2) (2 n+2 s+3) \ln \big[(n-k+s+1)(n+k+s+2)\big] \no \\
&\   \te -\frac{1}{12} (n+1) (2 k+1) (n+2 k+2) (2 n+2 k+3) \ln\big[(n+k-s+1)
   (n+k+s+2)\big]
    \no 
   \end{align}
 Here we introduced a   sharp    UV cutoff  $N$   for  the sum   over $n$  and 
did  not specify  the starting values of $n$ in the sum, which  should be different for  various terms
being related to  projection of zero modes: these   shifts   will  not   be relevant   for 
 the calculation of the  most trancendental  terms  in $G_s$.  We  made a particular 
   choice of the upper limit  as $\L-s-2$ 
 to  reproduce the  already   known (from proper time cutoff computation  \ci{Tseytlin:2013jya})  
  value of the $\aa$-coefficient  \rf{229}   of the logarithmic divergence. 

For example, in the    $s=1$ case  we obtain  from  (\ref{D.22})
\ba\no 
G_{1} &=\te  { 1 \ov 24} \L^4 \big( 4{\ln \L} -1 \big)+\frac{1}{3} \L^3 \ln 
   \L   - { 1 \ov 36} \L^2 \big({60 \ln \L} - 17 \big)
    +{1 \ov 6} \L
   \big( {\ln \L}-  1 \big)  +\frac{31 }{45} \ln 
   \L \no \\
   &\te  \ \ \ \  -\frac{11 }{3} \ln 
   A-\frac{2}{3} \zeta '(-3)-\frac{52}{135}  -{9\ov 2} \ln 3-5 \ln 
   2 +\mc O(\L^{-1}) \ ,\la{aa22}
\end{align}
where $A$ is Glaisher constant. 
$G_{s} $  turns out to  have  a  similar  structure  also for  $s>1$, i.e. one finds 
\ba
\no 
G_{s} &=  A_s(\L)  +  4 \aa_s \ln \L    +  F_s   +  \mc O(\L^{-1})   \   ,   \ \ \ \ \ \ \ \ 
\aa_s = \te \frac{1}{720}s^{2}(s+1)^{2}(14s^{2}+14s+3) \ ,  \\
 F_s & = \te   -\frac{1}{6}s(s+1)(5s^{2}+5s+1)\,\ln A-\frac{1}{3}s(s+1)\,\zeta'(-3)
+\dots \ , \label{D.24}
\end{align} 
where  $ A_s(\L)$  denote  all divergent  terms with positive powers of $\L$  and 
 $\aa_s$ is the same  conformal anomaly coefficient as in  \rf{229}\foot{To recall normalizations (see, e.g., \ci{Tseytlin:2013fca}), 
 $\ln Z= B_4 \ln L + ... $ where $L$ is a  UV cutoff and $B_4 = { 1 \ov (4 \pi)^2} \int d^4 x \sqrt g\,    b_4\Big|_{S^4} = - 4 \aa$, 
 where  as in \rf{333} we have $b_4 = - \aa R^*R^* + \cc C^2$. }
  and dots  stand for   rational  numbers  plus   logarithms of  integers. 
  
  Let us now turn to the $AdS_5$ side. 
 Each of the  MHS partition functions in the r.h.s. of \rf{20}  is given by the ratio of the physical and ghost  determinants as in \rf{9}
 with the   operators    $\hDelta_{s\, \perp}$      and  $\hDelta_{s-1\, \perp}$  
 having  the   "mass"   terms  $m^2 = \Delta(\Delta -d) - s$ 
 corresponding to dimensions $\Delta \equiv \Delta_+ = d+ s-2 $ and $d+ s-1$  respectively.
 Using the key  relation    (\ref{D.9})  and integrating over $\Delta$ 
  we can first  find the pole parts  in the limit $d=4-\va \to 4$:\foot{An alternative computation of these IR singular terms in  these $AdS_5$ partition functions 
  was  first done in \ci{Giombi:2013yva}.}
\ba
&\ln \frac{\det_-\hDelta_{s\, \perp}}{\det_+\hDelta_{s\, \perp}}\Big|_{\Delta= d+ s -2} = 
  {\te \frac{1}{90}\,s^{3}(s+1)^{2}(2s^{2}+10s+5)}  { 1\ov \va} +\dots. \ , \la{d23} \\
&
\ln \frac{\det_-\hDelta_{s-1\, \perp}}{\det_+\hDelta_{s-1\, \perp}}\Big|_{\Delta= d+ s -1}={\te 
\frac{1}{90}\,s^{2}(s+1)^{3}(2s^{2}-6s-3)} \frac{1}{\varepsilon} +... \la{d24}
\end{align}
Defining 
\be \la{d244}
\GG_s \equiv  -   \ln \frac{Z_{\MHS,s }^{-}}{Z_{\MHS,s }^{+}}
= \ha  \Big(\ln \frac{\det_-\hDelta_{s\, \perp}}{\det_+\hDelta_{s\, \perp}}\Big|_{\Delta= d+ s -2} 
- \ln \frac{\det_-\hDelta_{s-1\, \perp}}{\det_+\hDelta_{s-1\, \perp}}\Big|_{\Delta= d+ s -1}\Big) 
\ee 
and comparing to \rf{aa1},\rf{D.24}  we  check  that   the logarithmically divergent terms in the l.h.s.  and r.h.s. parts of \rf{20}  match 
(cf. \rf{D.11}) 
\be\la{d25}
 \big(G_{s}\big)_{\ln \L} =   \big(\GG_s\big)_{ 1\ov \va} \ . 
\ee
The finite part  of $\GG_s$  can be computed by  integrating the finite part of (\ref{D.9}),
separately for the physical and  the ghost contributions. As in the  spin  zero case   discussed  in 
the previous subsection, there are poles along the integration interval that  should be related to  the zero modes  on the CHS side, i.e. to
the  zero  eigenvalue contributions  in (\ref{D.22}) that should be projected out.
 The treatment  of these poles  is completely  analogous  to the one    in the scalar case  discussed below \rf{d5}. 
 We  find 
 \ba\no 
&   \ha  \Big(\ln \frac{\det_-\hDelta_{s\, \perp}}{\det_+\hDelta_{s\, \perp}}\Big|_{\Delta= d+ s -2-\delta} 
- \ln \frac{\det_-\hDelta_{s-1\, \perp}}{\det_+\hDelta_{s-1\, \perp}}\Big|_{\Delta= d+ s -1-\delta}\Big)_{\rm fin} \\
&\te  =  -\frac{1}{6} \left(4 s^3+6 s^2-1\right) \ln A-\frac{1}{12}  s^2 (s+1)^2
   (2 s+1) \ln\big[\Gamma(s)\Gamma(s+1)\big] \no \\
   &\quad\te  -\frac{1}{90}  s^2 (s+1)^2 \left(14 s^2+14 s+3\right) \gamma_{\rm E} 
   +\frac{1}{12} s 
   (2 s^4+5 s^3+2 s^2-2 s-1)  \ln (2\pi )\no \\
&\quad\te 
-\frac{1}{3}
   (2 s+1) \zeta '(-3)+\frac{1}{8 \pi ^2} (2 s+1) \zeta (3)
   +2 s^2 \psi ^{(-5)}(-s)+2  s^2(s+1) \psi ^{(-4)}(-s)\no \\
   &\quad\te +\frac{1}{6}
   \left(5 s^2+12 s+6\right) s^2 \psi ^{(-3)}(-s)+\frac{1}{180}
   (s+1)^3 \left(2 s^2-6 s-3\right) s^2 \psi ^{(0)}(s+1)\no \\
   &\quad\te 
   -\frac{1}{6}
   \left(5 s^4+8 s^3-4 s-1\right) \psi ^{(-3)}(1-s)-\frac{1}{180}
   (s+1)^2  s^3 \left(2 s^2+10 s+5\right) \psi ^{(0)}(s+2)\no\\
   &\quad\te 
   +\frac{1}{6}
   \left(s^3+5 s^2+6 s+2\right) s^2 \psi ^{(-2)}(-s)
   +\frac{1}{6}
   \left(-s^5+4 s^3+4 s^2+s\right) \psi ^{(-2)}(1-s)   \la{88d}\\
   &\quad\te 
   -2s (s+1)^2 
   \psi ^{(-4)}(1-s)-2 (s+1)^2 \psi ^{(-5)}(1-s)\no \\
   &\quad \te 
   +\frac{1}{4320} (1204
   s^6+2478 s^5+1035 s^4-620 s^3-417 s^2-80 s-22)-\frac{1}{12}  s^2 (s+1)^2 (2 s+1)
   \,\ln\delta  \no 
\end{align}
One can check that the coefficient of $\ln\delta$ term is indeed  the sum of multiplicities of the zero eigenvalues. 

 Dropping $\ln \delta$-terms, i.e. concentrating    on the remaining finite contribution   analogous to 
 (\ref{D.21}),   its   most transcendental part   can be  put into the following simple form\foot{Notice that like in \rf{D.21} 
   the coefficient of $  \zeta '(-3)$ 
 happens to be   proportional   to the number of dynamical degrees of freedom.}
\ba &\qquad \qquad 
 (\GG_s)_{\rm fin} = \te   -\frac{1}{6} q_s \ln A-\frac{1}{3}  \nu_s  \zeta '(-3)+\dots \ ,\la{d222}\\
& \ \ \ \ q_s = s (s+1)\big(5 s^2 + 5s +1\big)= \nu_s ( 5 \nu_s + 1)  \ , \ \ \ \ \ \ \  \qquad    \nu_s = s(s+1)   \ , \la{d26}
\end{align}
where   dots  stand  again  for a rational contribution plus a string of logarithms of primes plus a  $\gamma_{\rm E}$
term    that may be combined with the $1\ov \va$   pole term as in \rf{d6}. 
Comparing with (\ref{D.24}), we  conclude that the  most transcendental terms in finite parts of 
$G_s $ \rf{aa1} and $\GG_s$ \rf{d244} match, i.e.  in addition to \rf{d25} we  get 
\be\la{d255}
 \big(G_{s} \big)_{ \ln A,\  \zeta'(-3)   } =   \big(\GG_s\big)_{ \ln A,\  \zeta'(-3)} \ . 
\ee
This provides  a  non-trivial confirmation of the relation \rf{20}.

Finally, let  us   sum over all spins to  provide a check of the $(Z_\CHS)_\tot =1$ relation \rf{21}
using  the same summation prescription  \rf{16}  that implies  the vanishing of the total number of  dynamical degrees of freedom
of the CHS   theory and   the total   value of the     conformal anomaly coefficient $\aa$  in \rf{229},\rf{D.24}
 \ci{Giombi:2013yva,Giombi:2014iua}
\be \la{996}
 \n_\tot =\sum_{s=0}^\infty   \nu_s  \ e^{ -\ep (s + { 1 \ov 2})} \Big|_{\rm fin.} = 0 \ , \ \ \ \ \qquad  
 \aa_\tot  =\sum_{s=0}^\infty   \aa_s  \ e^{ -\ep (s + { 1 \ov 2})} \Big|_{\rm fin.} = 0 \ .
  \ee 
 We observe  that the same is true   also  for the sum of $q_s $   coefficients   in \rf{d26}\foot{Explicitly, 
 $\sum_{s=0}^{\infty}   s (s+1)(5 s^2+5 s+1) \, e^{-\epsilon(s+\frac{1}{2})}  
 = \frac{120}{\epsilon ^5}-\frac{3}{\epsilon ^3}+\frac{1}{16 \epsilon }+  {\cal O}( \epsilon) $.
 More generally, $ \sum_{s=0}^\infty  ( \nu_s)^n  \ e^{ -\ep (s + { 1 \ov 2})} \Big|_{\rm fin.} = 0$
  for $\nu_s = s(s+1)$ and any integer $n$.} 
 \be 
q_\tot  =\sum_{s=0}^\infty   q_s  \ e^{ -\ep (s + { 1 \ov 2})} \Big|_{\rm fin.} = 0 \ .  \la{699}
  \ee 
This   implies the vanishing   not only of the UV singular part  but also of   the most transcendental  finite part    
of  $\ln (Z_\CHS)_\tot$,   i.e.  of \rf{d222} summed over all spins.

\section{Conditions for  scalar gauge invariance of conformal  symmetric rank  $2$  tensor  in curved background }
\label{A:gauge}

Let us consider the variation of the Lagrangian (\ref{1.10}) under the transformation (\ref{1.11}). 
Integrating by parts in the  linear in $\sigma$  terms (moving covariant derivatives from $\sigma$ to the background and $\varphi_{\mu\nu}$)
 the condition for invariance  may be written as 
\be
\label{2.1}
\begin{split}
& -2 \na^{\mu}R \na_{\nu}\varphi_{\mu}{}^{\nu} -  R \na_{\nu}\na_{\mu}\varphi^{\mu \nu} -  
\varphi^{\mu \nu} \na_{\nu}\na_{\mu}R - 6 \na_{\nu}\na_{\mu}\na_{\rho}\na^{\rho}\varphi^{\mu 
\nu} + 12 \varphi^{\mu \nu} \na_{\nu}\na_{\rho}R_{\mu}{}^{\rho}  \\ 
& + 8 \na_{\nu}\na_{\rho}\na^{\rho}\na_{\mu}\varphi^{\mu \nu} + 12 \na_{\mu}\varphi^{\mu \nu} 
\na_{\rho}R_{\nu}{}^{\rho} + 12 R^{\mu \nu} \na_{\rho}\na_{\nu}
\varphi_{\mu}{}^{\rho} - 3 R^{\mu \nu} \na_{\rho}\na^{\rho}\varphi_{\mu \nu}  \\ 
& - 3 \varphi^{\mu \nu} \na_{\rho}\na^{\rho}R_{\mu \nu} - 2 \na_{\rho}\na^{\rho}
\na_{\nu}\na_{\mu}\varphi^{\mu \nu} + 12 \na_{\nu}R_{\mu \rho} \na^{\rho}\varphi^{\mu \nu} 
- 6 \na_{\rho}R_{\mu \nu} \na^{\rho}\varphi^{\mu \nu}\\ 
& + 6\omega (
\varphi^{\mu \nu} \na_{\lambda}\na_{\rho}C_{\mu}{}^{\rho}{}_{\nu}{}^{\lambda}
 +  C_{\mu \rho \nu \lambda} 
\na^{\lambda}\na^{\rho}\varphi^{\mu \nu} + 2 \na_{\lambda}C_{\mu \rho \nu}{}^{\lambda} \na^{\rho}\varphi^{\mu \nu}) = 0.
\end{split}
\ee
Commuting  the covariant derivatives in the $\na^{4} \varphi$ terms  we find that 
 they cancel  against  the $\na^{3}\varphi$ terms. The remaining $\na^{2}\varphi$ terms  take the form 
\be
 \tfrac{1}{3} R \na_{\nu}\na_{\mu}\varphi^{\mu \nu} 
-  \tfrac{4}{3} R^{\mu \nu} \na_{\nu}\na_{\rho}\varphi_{\mu}{}^{\rho} +R^{\mu \nu} \na_{\rho}\na^{\rho}
\varphi_{\mu \nu} + 2(2+\omega) C_{\mu \rho \nu \lambda} \na^{\lambda}\na^{\rho}\varphi^{\mu \nu} \ . 
\ee
Since  $\varphi_{\mu\nu}$   is  symmetric traceless   this gives the  condition 
$
\widetilde K_{\mu\nu\rho\sigma} = K_{\mu\nu\rho\lambda}-\frac{1}{4}g_{\mu\nu}\,K^{\sigma}{}_{\sigma\rho\lambda} = 0,
$
where 
\be \la{233} 
\begin{split}
K_{\mu\nu\rho\lambda}&= g_{\rho \lambda} R_{\mu \nu} -  \tfrac{2}{3} (g_{\nu \rho} R_{\mu \
\lambda} + g_{\mu \rho} R_{\nu \lambda}) + \tfrac{1}{6} (g_{\mu \rho} \
g_{\nu \lambda} + g_{\mu \lambda} g_{\nu \rho}) R \\
& \ \ \ \ + (2 + \omega) \
(C_{\mu \rho \nu \lambda} + C_{\nu \rho \mu \
\lambda}) = 0\ .
\end{split}
\ee
Then the contraction  $g^{\mu\rho}\widetilde K_{\mu\nu\rho\sigma}=0$ gives the  requirement that  the background  should be 
 Einstein
\be
\label{2.5}
R_{\mu\nu} =\te  \frac{1}{4} R \, g_{\mu\nu}\  .
\ee
Using this in  \rf{233}    and $
\widetilde K_{\mu\nu\rho\sigma} =0$  gives further  constraint 
\be
(2+\omega) C_{ \rho (\mu \nu)  \lambda}=0 \ .
\ee
If $2+\omega\not=0$  then  $C_{ \rho (\mu \nu)  \lambda}=0 $   combined 
with  the first Bianchi identity
$C_{\mu[\nu\rho\lambda]} = C_{\mu\nu\rho\lambda}+C_{\mu\rho\lambda\nu}+C_{\mu\lambda\nu\rho}=0,
$
implies that $C_{\mu\nu\rho\lambda}=0$, i.e.   the space  should be conformally flat. 
The alternative   is  to assume that 
\be  \omega=-2 \ . \la{255} \ee 
Then 
 the remaining part of the  variation  \rf{2.1}  gives  the  condition
$ \na_{\lambda}C^{\lambda}_{(\mu \nu) \rho}{} = 0 \ ,  $  {i.e.} $ \  \na_{\lambda}C^{\lambda}{}_{\mu\nu\rho} =0.
$ 
This is  automatically   satisfied  as a  consequence of the 
 the Einstein condition (\ref{2.5})  and the second Bianchi identity $\na_{[\lambda}R_{\mu\nu]\sigma  \rho}=0$. 
 
 In conclusion, the Lagrangian (\ref{1.10}) admits the invariance (\ref{1.11}) if the background is Einstein  
 and  is  also conformally flat  or  it is generic  but  then  $\omega $ is to be fixed as   in  \rf{255}. 
 
 A similar   analysis for $s >2$   implies  that   imposing the Einstein condition \rf{2.5}  is not enough
 to ensure  the   scalar   gauge invariance of \rf{1.7} for any value of $\omega$  unless the space is also  conformally flat.


\section{Partition function    of conformal symmetric  rank 2 tensor on $S^{1}\times S^{3}$  
}
\label{B:k}

Starting  with the rank 2 tensor Lagrangian \rf{1.10} on conformally-flat $S^1_\beta  \times S^3$   space and 
performing 1+3 decomposition $\varphi_{\mu\nu} = (\varphi_{ij}, \varphi_{0i}, \varphi_{00})$, 
$\varphi_{ij} = \varphi_{ij}^{\perp}+\na_{(i}V^{\perp}_{j)}+\dots$ one  can represent  the resulting partition function as 
(cf. \rf{2.21}) 
 \be
\label{2.23}
Z _{\conf,2} = \Big[\frac{1}{\det {\bDelta}_{2\, \perp}\, \det'{\bDelta}_{1\, \perp} \det{\bDelta}_{1\, \perp}   }\Big]^{1/2} \ , \ee
where $\bDelta_n $ operators act on 3d  $n$-tensors. We consider  unit-radius $S^3$  and  $S^1$ of length $\beta$.
Here 
${\bDelta}_{2\, \perp}=  -\na^{2}+3 = -\partial_{0}^{2}-\mathbf{\na}^{2}+3$
and  the two  vector operators   acting on $V^\perp_i$ and $\vp_{0i}$     have   similar   form  as in Maxwell theory 
(cf. \ci{Beccaria:2014jxa}).
The spectrum  of the rank 2 operator is found to be 
\be
\lambda_{k,n}= ( {2 \pi k \ov \beta})^2 + w_n^2  \ , \ \ \ \ \ \   w^{2}_n=
(n+2)(n+4)-2+3 =(n+3)^{2}  \ . 
\ee
As a result,\ \   $\ln  Z _{\conf,2} =  \sum_{m=1}^\infty { 1 \ov m}  \Z (q^m)$, 
where $q= e^{-\beta} $   and  the one-particle partition function $\Z(q)$ is 
\ba
\Z(q) 
&= \underbrace{\sum_{n=0}^{\infty}2\,(n+1)(n+5)q^{n+3}}_{\varphi_{ij}^{\perp}}+
\underbrace{\sum_{n=1}^{\infty}2\,(n+1)(n+3)q^{n+1}}_{V_{i}^{\perp}}+
\underbrace{\sum_{n=0}^{\infty}2\,(n+1)(n+3)q^{n+3}}_{\varphi_{0i}} \no \\
&=  \frac{2(8\,q^{2}-9\,q^{3}+\,q^{5})}{(1-q)^{4}} 
  \ . \la{2.22} 
\end{align}
Here the $V^\perp_i$    contribution  starts at $n=1$ because, as in  \cite{Beccaria:2014jxa}, the 6 zero modes of this vector  drop out. 

The same   expression can be found by counting  the conformal operators in flat space $\mathbb R^4$ (cf.  \ci{Kutasov:2000td, Beccaria:2014jxa}).
The flat-space equations  \rf{1.2} may be written in terms  of a field  strength invariant under the  scalar gauge transformations \ci{Anselmi:1999bb} 
\ba
\label{1.1.9}\te 
H_{\mu\nu\rho} = \partial_{[\mu}\vp_{\nu]\rho}-\frac{1}{3}\delta_{\rho[\m}\partial_{\alpha}
\vp^{\alpha}_{\nu]} \ ,  \qquad \qquad  H_{\mu\nu}^{\ \ \ \nu}=0, \qquad  \epsilon^{\alpha\mu\nu\rho}H_{\mu\nu\rho}=0.
\end{align}
The number of independent components of  dimension 2 field $H_{\mu\nu\rho}$ is   $6\times 4-4-4=16$.
The equations of motion  together  with "Bianchi" identities then take the form 
\be
\la{118} \te 
\partial^{\mu}\,  H_{\mu (\nu\rho)}=0 \  , \ \ \ \qquad 
\partial^{\mu}\,\widetilde H_{\mu(\nu\rho)}=0 \ , \qquad \qquad 
\widetilde H_{\mu\nu\rho} \equiv  \frac{1}{2}\epsilon_{\mu\nu\alpha\beta}\,H^{\alpha\beta}_{\ \ \ \rho}\ , \ \ \ \ \ \ \ \ 
\widetilde H_{\mu\nu}^{\ \ \ \nu} = 0 \ ,  
\ee
which  are  symmetric under  $H \leftrightarrow \widetilde H$.
{An explicit count} of all gauge invariant operators $\del ... \del H$  modulo equations of motion  and identities gives\footnote{
Note that   $\Z _{\conf,2}= \Z_2 + \Z_1 $   where  $\Z_{1} = {2 q^2 ( 3-q)\ov (1-q)^3}$  and 
$\Z_2 ={ 2 q^2 ( 5-q^2)\ov (1-q)^3}$. Here   $\Z_{s}= {2 q^2 ( n_{s} -q^s)\ov (1-q)^3}$
where $n_{s}$ is the number of  physical off-shell  d.o.f. (number of components  minus gauge parameters).
}
 \be \Z _{\conf,2}(q) =  \frac{2q^2 (8-\,q-\,q^{2})}{(1-q)^{3}} \ , \la{345} 
 \ee 
 which is   equivalent to  \rf{2.22}. 

\section{Partition function  of conformal  symmetric rank 3  tensor   on $S^{4}$}

The Lagrangian (\ref{1.7}) for $s=3$    on unit-radius $S^4$  background    is 
\be\la{la3}\te 
\mathscr L_{\conf,3}(S^4)= \na^{\lambda}\varphi^{\mu\nu\rho}\,\na_{\lambda}\varphi_{\mu\nu\rho}-\frac{3}{2}
\na_{\rho}\varphi^{\mu\nu\rho}\,\na^{\lambda}\varphi_{\mu\nu\lambda}
+5\,\varphi^{\mu\nu\rho}\varphi_{\mu\nu\rho}.
\ee
Decomposing $\varphi_{\mu\nu\rho}$ as\foot{Note that  the  scalar curvature  here $R=12$.}
\be\la{tra}
\begin{split}
\varphi_{\mu\nu\rho} = &\varphi_{\mu\nu\rho}^{\perp}+\na_{(\mu}h_{\nu\rho)}^{\perp}
+\na_{(\mu}\na_{\nu}V_{\rho)}^{\perp}+\na_{(\mu}\na_{\nu}\na_{\rho)}\sigma\\
&\te -\frac{1}{2}g_{(\mu\nu}V_{\rho)}^{\perp}
-\frac{1}{6}g_{(\mu\nu}\na^{2}V^{\perp}_{\rho)}-g_{(\mu\nu}\na_{\rho)}\sigma
-\frac{1}{2} g_{(\mu\nu}\na_{\rho)}\na^{2}\sigma,
\end{split}
\ee
we get 
\be\la{laa}
\begin{split}
\mathscr L_{\conf,3} = &\te 
\varphi_{\mu\nu\rho}^{\perp}\Delta_{3\,\perp}(5)\varphi^{\perp\,\mu\nu\rho}+
\frac{1}{6}\,h_{\mu\nu}^{\perp}\Delta_{2\,\perp}(-8)\, \Delta_{2\,\perp}(4)\,h^{\perp\,\mu\nu}\\
&\te 
+\frac{5}{108}V_{\mu}^{\perp}\Delta_{1\,\perp}(-9)\, \Delta_{1\,\perp}(-3)\, \Delta_{1\,\perp}(3)\,V^{\perp\,\mu},
\end{split}
\ee
where $\sigma$ decouples due to scalar gauge invariance.
Here $\Delta(M^2) = - \na^2 + M^2$ as   in \rf{7}. 
The   Jacobian of transformation \rf{tra}   can be found from 
\be\la{ja}
\begin{split}
&\te \int d^{4}x \sqrt{g}\, \varphi^{\mu\nu\rho}\varphi_{\mu\nu\rho} = 
\int d^{4}x \sqrt{g}\, \Big[ 
\varphi_{\mu\nu\rho}^{\perp}\varphi^{\perp\,\mu\nu\rho}
+\frac{1}{3}\,h_{\mu\nu}^{\perp}\Delta_{2\,\perp}(-8)\, h^{\perp\,\mu\nu}\\
&\te \ \ \ +\frac{5}{18}V_{\mu}^{\perp}\Delta_{1\,\perp}(-9)\Delta_{1\,\perp}(-3)V^{\perp\,\mu}
+\frac{1}{2}\sigma\,\Delta_{0}(0)\,\Delta_{0}(-4)\,\Delta_{0}(-10)\sigma\Big].
\end{split}
\ee
The resulting partition function is thus (cf. \rf{2.17},\rf{217}) 
\be\la{z3}
Z_{\conf,3} = \Big[
\frac{\det\Delta_{0}(0)\det\Delta_{0}(-4)\det\Delta_{0}(-10)}{
\det\Delta_{3\perp}(5)\det\Delta_{2\perp}(4)\det\Delta_{1\perp}(3)}\Big]^{1/2} \ .
\ee

\bibliography{CST-Biblio}

\bibliographystyle{JHEP}

\end{document}